\documentclass[useAMS,usenatbib]{mnras}
\usepackage{natbib}
\usepackage{amsmath}
\usepackage{graphicx}
\usepackage{multirow}
\usepackage{url}
\numberwithin{equation}{section}
\title[Radiative cooling in AGN winds]{Radiative cooling of swept up gas in AGN-driven galactic winds and its implications for molecular outflows}
\author[A. J. Richings and C.-A. Faucher-Gigu\`{e}re]{Alexander J. Richings$^{1}$\thanks{Email: a.j.richings@northwestern.edu} and Claude-Andr\'{e} Faucher-Gigu\`{e}re$^{1}$\\
$^{1}$Center for Interdisciplinary Exploration and Research in Astrophysics (CIERA) and Department of Physics and Astronomy,\\ 
Northwestern University, 2145 Sheridan Road, Evanston, IL 60208, USA}

\begin{document}

\date{\today}

\pagerange{\pageref{firstpage}--\pageref{lastpage}} \pubyear{2017}

\maketitle

\label{firstpage}

\begin{abstract} 
We recently used hydro-chemical simulations to demonstrate that molecular outflows observed in luminous quasars can be explained by molecule formation within the AGN wind. However, these simulations cover a limited parameter space, due to their computational cost. We have therefore developed an analytic model to follow cooling in the shocked ISM layer of an AGN wind. We explore different ambient densities (1$-$10$^{4} \, \rm{cm}^{-3}$), density profile slopes (0$-$1.5), AGN luminosities (10$^{44}$$-$10$^{47} \, \rm{erg} \, \rm{s}^{-1}$), and metallicities (0.1$-$3 Z$_{\odot}$). The swept up gas mostly cools within $\sim$1 Myr. Based on our previous simulations, we predict that this gas would produce observable molecular outflows. The instantaneous momentum boost initially increases as the outflow decelerates. However, it reaches a maximum of $\approx$20, due to work done against the gravitational potential. The predicted time-averaged observational estimate of the molecular outflow momentum boost reaches a maximum of $\approx$1$-$2, partly due to our assumed molecular fraction, 0.2, but also because the instantaneous and observational, time-averaged definitions are not equivalent. Thus recent observational estimates of order unity momentum boosts do not necessarily rule out energy-driven outflows. Finally, we find that dust grains are likely to re-form by accretion of metals after the shocked ISM layer has cooled, assuming that a small fraction of dust grains swept up after this layer has cooled are able to mix into the cool phase, and assuming that grain growth remains efficient in the presence of the strong AGN radiation field. This would enable rapid molecule formation, as assumed in our models.
\end{abstract}


\section{Introduction}\label{intro_sect} 

There is a wealth of observational evidence for multiphase outflows on galactic scales, which have been seen in ionized \citep[e.g.][]{heckman90, greene12, harrison12, liu13}, neutral atomic \citep[e.g.][]{martin05,rupke05,rupke11} and molecular gas \citep[e.g.][]{aalto12, cicone12, cicone14, feruglio13a, feruglio17, gonzalezalfonso17}; see also \citet{rupke17}, who present observations of multiphase winds in type 1 quasars. These outflows can be driven by active galactic nuclei (AGN) and/or by star formation in the galaxy. 

The presence of cool ($\la 10^{4} \, \rm{K}$) gas in these outflows may at first seem surprising, given that their high velocities ($\sim 100 - 1000 \, \rm{km} \, \rm{s}^{-1}$) suggest post-shock temperatures $\sim 10^{5} - 10^{7} \, \rm{K}$. One possible explanation is that the cool outflowing gas originates from cool clouds that were swept up from the ISM of the host galaxy by the hot wind, and were accelerated by ram pressure \citep[e.g.][]{veilleux05}. \citet{gaspari17} present a model for AGN feeding and feedback in which they argue that an energy-driven outflow will entrain hot ($\sim$10$^{7} \, \rm{K}$), warm ($\sim$10$^{4 - 5} \, \rm{K}$) and cold ($\la$100 K) gas, producing multiphase outflows with velocities and outflow rates in good agreement with observations. However, cool clouds that are accelerated by a hot, fast wind are likely to be rapidly destroyed by hydrodynamical instabilities \citep{klein94, scannapieco15, bruggen16, schneider17, zhang17}. Other mechanisms may also accelerate cool gas from the host galaxy, such as acceleration by cosmic rays \citep{socrates08, booth13, hanasz13, simpson16} or radiation pressure acting on dust grains \citep{murray11, krumholz13, thompson15}. 

Alternatively, rather than sweeping up existing cool gas from the host galaxy, the hot wind may itself cool and form cold clumps within the outflow \citep{wang95, silich03, martin15, scannapieco17}. \citet{zubovas14} explored gas cooling in a spherically symmetric AGN outflow in the energy-driven regime, based on the outflow models of \citet{king05} and \citet{zubovas12}. They showed that thermal instabilities in the outflow would lead to a two-phase medium, and argued that the cool gas would become molecular and form stars. 

\citet{thompson16} also explored gas cooling in galactic winds. Their study focussed on starburst-driven winds, using the steady-state wind model of \citet{chevalier85}, in which energy and mass are injected at a constant rate within a finite radius that represents the starburst region. They showed that the wind will radiatively cool as long as the mass loading is sufficiently high. 

In \citet[][hereafter Paper~\textsc{i}]{richings18}, we ran a series of hydrodynamic simulations of an isotropic AGN wind interacting with a uniform ambient medium, including a time-dependent model for the chemistry of ions and molecules, to investigate the origin of fast molecular outflows in quasars. In these simulations, the small-scale wind was modelled after the properties of accretion disc winds observed as broad absorption lines (BALs) in the UV \citep[e.g.][]{weymann81, gibson09} or ultra-fast outflows (UFOs) in X-rays \citep[e.g.][]{feruglio15, nardini15, tombesi15}, with velocities $\sim 0.1 c$. We showed that, in these simulations, the gas swept up from the ambient medium by the outflow was able to cool and form molecules within $\sim 1 \, \rm{Myr}$, producing molecular outflow rates up to $140 \, \rm{M}_{\odot} \, \rm{yr}^{-1}$. Thus in-situ molecule formation can potentially account for observed molecular outflow rates in quasars, for ambient densities $\ga 10 \, \rm{cm}^{-3}$ and metallicities of at least solar. However, the high computational cost of these simulations meant that we could only consider a limited range of ambient densities ($1 - 10 \, \rm{cm}^{-3}$), AGN luminosities ($10^{45} - 10^{46} \, \rm{erg} \, \rm{s}^{-1}$), and metallicities ($0.1 - 1 \, \rm{Z}_{\odot}$). 

To explore the feasibility of molecule formation across a wider range of physical parameters relevant to AGN host galaxies, we can make use of the analytic model of \citet[][hereafter FGQ12]{fauchergiguere12}, which generalised classic stellar wind models \citep[e.g.][]{weaver77, koo92}. Similar models were previously studied in the AGN context by e.g. \citet{king11}, but with different results. The FGQ12 model follows the evolution of a spherically symmetric outflowing shell driven by a central AGN as it sweeps through the ambient medium. This is almost identical to the set up of the simulations in Paper~\textsc{i}, except that, in FGQ12, the ambient medium has a power-law density profile with radius, whereas in our simulations the ambient medium was uniform, i.e. the density profile slope was zero. In the FGQ12 model, the AGN launches a fast ($\sim 30 \, 000 \, \rm{km} \, \rm{s}^{-1}$) wind on small (sub-pc) scales. This wind shocks, creating a hot bubble that drives an outflow into the ambient medium. FGQ12 showed that, for physical conditions typical of quasar host galaxies, cooling in the hot wind bubble is inefficient, which results in an energy-driven outflow. Note that, unlike the \citet{chevalier85} model, the FGQ12 model is not a steady-state wind solution. Instead, it follows the time-dependent evolution of a single outflowing shell as it propagates outwards into the ambient medium. 

\citet{nims15} used the FGQ12 model to predict the observable emission from AGN winds. They noted that, while cooling in the hot wind bubble is inefficient, the shell of gas swept up from the ambient medium (i.e. the shocked ISM layer) can cool quickly, due to the higher densities in this layer (a point verified by our simulations in Paper~\textsc{i}). They thus demonstrated that the radiatively cooling shocked ISM layer will produce strong emission due to free-free and inverse Compton cooling (in X-rays) and synchrotron emission (from radio to X-rays). 

In this paper, we extend the FGQ12 model to explicitly follow radiative cooling in the shocked ISM layer, down to $10^{4} \, \rm{K}$. Below this temperature, we would need to follow the molecular chemistry, to account for molecular cooling. Furthermore, we saw in Paper~\textsc{i} that, once the shocked ISM layer has cooled below $\sim 10^{4} \, \rm{K}$, it forms a complex multiphase medium, and can no longer be reasonably represented by a single density and temperature. However, we also showed in Paper~\textsc{i} that, once the shocked ISM layer has cooled below $10^{4} \, \rm{K}$, it continues to cool quickly to even lower temperatures ($<10^{3} \, \rm{K}$) where molecules can form rapidly, assuming a Milky Way dust-to-metals ratio. We can therefore use this analytic model to predict when the outflow cools to $10^{4} \, \rm{K}$, and then, based on the results of our simulations from Paper~\textsc{i}, we predict that the outflow will become molecular once it has cooled below this temperature. We can then use this model to predict molecular outflow rates and momentum boost factors, which we compare to observations. The analytic model thus enables us to test the predictions of our molecular outflow simulations against observations across a wider range of physical parameters than with the simulations alone. We apply this model to a wide range of ambient densities ($1 \leq n_{\rm{H}} \leq 10^{4} \, \rm{cm}^{-3}$), density profile slopes ($0 \leq \alpha \leq 1.5$), AGN luminosities ($10^{44} \leq L_{\rm{AGN}} \leq 10^{47} \, \rm{erg} \, \rm{s}^{-1}$), and metallicities ($0.1 \leq Z \leq 3 \, \rm{Z}_{\odot}$). 

Unlike previous studies that also considered cooling of the swept up shell in similar analytic models \citep[e.g.][]{zubovas14, wang15}, this is the first study that tests the analytic model systematically by comparing it to hydrodynamic simulations. It is also the first time that hydro-chemical simulations have been used to calibrate the efficiency of molecule formation. 

The remainder of this paper is organised as follows. We briefly summarise the details of our simulations from Paper~\textsc{i} in Section~\ref{sims_sect}, and we describe the analytic model of FGQ12, along with our modifications to this model, in Section~\ref{analytic_model_sect}. In Section~\ref{comparison_sect} we compare the simulations from Paper~\textsc{i} to the analytic model. We use the analytic model to explore a wide range of parameters in Section~\ref{param_sect}, and we compare the analytic model to observations in Section~\ref{obs_sect}. In Section~\ref{dust_sect}, we present a model for dust formation and destruction within the shocked ISM layer, which we implement in our AGN wind model. We discuss our conclusions in Section~\ref{conclusions_sect}. We present resolution tests in Appendix~\ref{resolution_appendix}, and we derive an approximate prescription for inverse Compton cooling of a hot shocked wind, taking into account two-temperature plasma effects, in Appendix~\ref{2T_cool_sect}. 

\section{Simulations}\label{sims_sect} 

In Paper \textsc{i} we presented a series of hydrodynamic simulations of an isotropic AGN wind interacting with a uniform ambient medium. These simulations included a treatment for the time-dependent chemistry, in particular to follow the formation and destruction of molecules. 

The simulations were run with the \textsc{gizmo} code, using the Meshless Finite Mass (MFM) Lagrangian hydrodynamics method \citep{hopkins15}. Chemical abundances and radiative cooling were evolved in the simulations using the \textsc{chimes} chemistry module \citep{richings14a, richings14b}, which follows the evolution of 157 species, including all ionization states of 11 elements that are important for cooling, and 20 molecules, most notably H$_{2}$, CO and OH. The \textsc{chimes} module then calculates cooling and heating rates from the non-equilibrium chemical abundances, and integrates the temperature in time along with the 157 rate equations. 

The chemistry network includes collisional ionization, photoionization, cosmic ray ionization, recombination (both radiative and di-electronic), charge transfer, formation of H$_{2}$ on dust grains and in the gas phase, and other molecular reactions. A full list of the chemical reactions in \textsc{chimes} can be found in table~B1 of \citet{richings14a}. Thermal processes include atomic line cooling from H, He and metals, molecular cooling from H$_{2}$, CO, H$_{2}$O and OH, bremsstrahlung cooling, non-relativistic Compton cooling/heating from the AGN radiation field, photoheating, photoelectric heating from dust grains, and cosmic ray heating (see table~1 of \citealt{richings14a}). 

To calculate the photochemical rates, we used the average quasar spectrum from \citet{sazonov04}. The gas was shielded from the radiation field using a local approximation, where we calculated a local shielding length using a Sobolev-like approximation as: 

\begin{equation}\label{sobolev_eqn} 
L_{\rm{sh}} = \frac{1}{2} \left( h_{\rm{inter}} + \frac{\rho}{|{\nabla \rho}|} \right), 
\end{equation} 
where $\rho$ is the density of the particle, and the inter-particle spacing is defined as $h_{\rm{inter}} = (m_{\rm{gas}} / \rho)^{1/3}$ for a particle with mass $m_{\rm{gas}}$. The resulting column density, $N_{\rm{H_{tot}}} = n_{\rm{H_{tot}}} L_{\rm{sh}}$ (where $n_{\rm{H_{tot}}}$ is the hydrogen number density), was then used to attenuate the photochemical rates. 

We assumed that the dust abundance scales linearly with metallicity. However, it is not clear whether dust grains can survive the strong shocks and high gas temperatures in AGN winds. For example, \citet{ferrara16} showed that, in their numerical simulations of the shocked ISM layer of an AGN wind, dust grains were rapidly destroyed by sputtering (within $10^{4} \, \rm{yrs}$). The dust abundance is therefore a major uncertainty in our simulations, although, as we discussed in Paper \textsc{i}, it may still be possible for dust grains to re-form in the shocked ISM layer after it has cooled, for example by accretion of metals from the gas phase onto grains (which was not included in the simulations of \citealt{ferrara16}). In section~\ref{dust_sect}, we develop an analytic model of dust destruction and formation. We show that, for a wide range of conditions representative of those expected in AGN outflows, dust grains are likely to re-form via metal accretion once the shocked ISM layer has cooled, assuming that a small fraction ($\ga 10^{-6}$) of the dust grains swept up by the outflow after the shocked ISM layer has cooled can be mixed into the cool phase, and assuming that grain growth can proceed efficiently in the presence of the strong UV radiation from the AGN. This grain re-formation would thus enable rapid molecule formation, as assumed in our simulations from Paper~\textsc{i}. 

The simulations were performed in a 3D periodic box, $1.6 - 5.0 \, \rm{kpc}$ across (depending on the simulation), and were run for $1 \, \rm{Myr}$. This corresponds to the typical flow time ($t_{\rm{flow}} = r / v$) of observed molecular outflows in AGN host galaxies \citep[e.g.][]{gonzalezalfonso17}. One octant of the simulation box was set up to be at a higher resolution, with $30 \, \rm{M}_{\odot}$ per gas particle (and 32 kernel neighbours) for the fiducial runs, while the remainder of the box used 8 times lower mass resolution. 

Since we considered an ambient medium with a uniform density, these simulations are more representative of outflows before they break out of the galactic disc. After the outflow breaks out of the disc, the ambient density can drop quickly, and the swept up gas will coast rather than being pushed by an over-pressurised bubble. 

We included the gravitational potential from the central black hole, using a single collisionless particle with a mass of $M_{\rm{BH}} = 10^{8} \, \rm{M}_{\odot}$, and the host galaxy. We modelled the galaxy potential with an isothermal sphere, for which the mass enclosed within a radius $R$ is: 

\begin{equation}\label{mgal_potential_eqn} 
M_{\rm{gal}}(<R) = \frac{2 \sigma^{2} R}{G}. 
\end{equation} 
This profile is parameterised by the velocity dispersion, which we set to $\sigma = 200 \, \rm{km} \, \rm{s}^{-1}$. This corresponds to the $M_{\rm{BH}} - \sigma$ relation \citep[e.g.][]{gultekin09} for our adopted black hole mass. Self-gravity of the gas was also included. The gravitational softening of gas particles was set equal to their inter-particle spacing, $h_{\rm{inter}}$, down to a minimum of $0.1 \, \rm{pc}$ at our fiducial resolution. The gravitational softening of the black hole particle was $1 \, \rm{pc}$. 

To drive the AGN wind, we injected wind particles within the central $1 \, \rm{pc}$ with an outward velocity $v_{\rm{in}} = 30 \, 000 \, \rm{km} \, \rm{s}^{-1}$. This initial velocity is motivated by observations of broad absorption line (BAL) quasars \citep[e.g.][]{weymann81, gibson09}, and by X-ray observations of ultra-fast outflows in quasars \citep[e.g.][]{feruglio15, nardini15, tombesi15}. The momentum injection rate is determined from the AGN luminosity, $L_{\rm{AGN}}$, according to: 

\begin{equation}\label{momentum_inj_eqn} 
\dot{M}_{\rm{in}} v_{\rm{in}} = \tau_{\rm{in}} \frac{L_{\rm{AGN}}}{c}, 
\end{equation} 
where $\dot{M}_{\rm{in}}$ is the mass injection rate, and the parameter $\tau_{\rm{in}}$ is set to unity. 

The parameters of each simulation are summarised in table~1 of Paper~\textsc{i}. In this work, we will focus on the four parameter variation runs at the fiducial resolution. These cover a range of ambient ISM densities ($n_{\rm{H}0}$), AGN luminosities ($L_{\rm{AGN}}$), and metallicities ($Z$). In our fiducial model (nH10\_L46\_Z1), we set $n_{\rm{H}0} = 10 \, \rm{cm}^{-3}$, $L_{\rm{AGN}} = 10^{46} \, \rm{erg} \, \rm{s}^{-1}$, and $Z = \rm{Z}_{\odot}$. We then performed a low-density run (nH1\_L46\_Z1, with $n_{\rm{H}0} = 1 \, \rm{cm}^{-3}$), a low-luminosity run (nH10\_L45\_Z1, with $L_{\rm{AGN}} = 10^{45} \, \rm{erg} \, \rm{s}^{-1}$), and a low-metallicity run (nH10\_L46\_Z0.1, with $Z = 0.1 \, \rm{Z}_{\odot}$). 

\section{Analytic model}\label{analytic_model_sect} 

\begin{figure}
\centering
\mbox{
	\includegraphics[width=84mm]{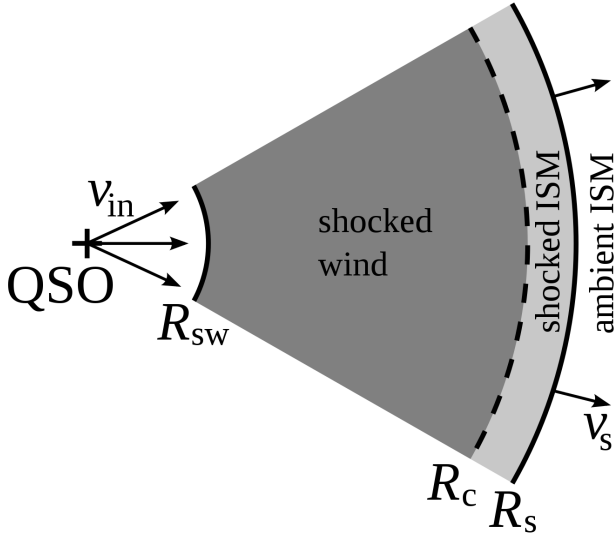}}
\caption{Schematic diagram of the AGN wind structure in the energy-driven regime. The QSO launches a small-scale ($\ll 1 \, \rm{pc}$) wind with velocity $v_{\rm{in}}$, which shocks at a radius $R_{\rm{sw}}$ (the reverse shock), creating a shocked wind bubble. A forward shock, at radius $R_{\rm{s}}$, propogates into the ambient ISM at a velocity $v_{\rm{s}}$, creating a layer of shocked ISM gas. This is separated from the shocked wind by a contact discontinuity, at radius $R_{\rm{c}}$. In this work, we focus on radiative cooling in the shocked ISM layer.} 
\label{windStructFig}
\end{figure}

The analytic model of FGQ12 considers an AGN with luminosity $L_{\rm{AGN}}$ that launches a wind with an initial velocity $v_{\rm{in}}$ on small scales ($\ll 1 \, \rm{pc}$). We take a fiducial value of $v_{\rm{in}} = 30 \, 000 \, \rm{km} \, \rm{s}^{-1}$, as in the simulations. The rate at which the AGN injects momentum into the wind is as given in equation~\ref{momentum_inj_eqn}, where we again take $\tau_{\rm{in}} = 1$. The wind material is shocked by a reverse shock at a radius $R_{\rm{sw}}$, creating a shocked wind bubble, while a forward shock propagates into the ambient ISM, creating a layer of shocked ISM material at a radius $R_{\rm{s}}$. A contact discontinuity, at radius $R_{\rm{c}} \approx R_{\rm{s}}$, separates these two regions. Fig.~\ref{windStructFig} shows a schematic diagram of the AGN wind structure. 

The FGQ12 model follows the evolution of the forward shock, $R_{\rm{s}}(t)$, with time $t$ by integrating the equation of motion (their equation 28): 

\begin{equation}\label{eqn_of_motion} 
\frac{\rm{d}}{\rm{d}\mathit{t}} (M_{\rm{s}} v_{\rm{s}}) = 4 \pi R_{\rm{s}}^{2} (P_{\rm{b}} - P_{0}) - \frac{G M_{\rm{s}} M_{\rm{t}}}{R_{\rm{s}}^{2}}, 
\end{equation} 
where $P_{\rm{b}}$ and $P_{0}$ are the thermal pressures of the shocked wind bubble and the ambient medium, respectively. We calculate $P_{0}$ assuming a constant ambient medium temperature of $10^{4} \, \rm{K}$, which we find in the simulations (see fig. 5 of Paper \textsc{i}). The shocked ISM layer has a velocity $v_{\rm{s}} = \dot{R}_{\rm{s}}$, and a mass $M_{\rm{s}} = \int_{0}^{R_{\rm{s}}} \rho_{\rm{g}} \rm{d}\mathit{V}$. The density profile of the ambient medium is taken to be a power law, $\rho_{\rm{g}} (R) = \rho_{0} (R / R_{0})^{-\alpha}$, which we normalise at $R_{0} = 100$ pc. 

The total gravitational mass within $R_{\rm{s}}$ is $M_{\rm{t}} = M_{\rm{BH}} + M_{\rm{gal}}(<R_{\rm{s}})$, where $M_{\rm{BH}}$ is the mass of the black hole and $M_{\rm{gal}}(<R_{\rm{s}})$ is the mass of the host galaxy within $R_{\rm{s}}$. We model the gravitational potential of the host galaxy as an isothermal sphere, as in the simulations, with an enclosed mass within a radius $R$ given by equation~\ref{mgal_potential_eqn}. We take fiducial values of $\sigma = 200 \, \rm{km} \, \rm{s}^{-1}$ and $M_{\rm{BH}} = 10^{8} \, \rm{M}_{\odot}$, as used in the simulations. 

To calculate $P_{\rm{b}}$, the FGQ12 model follows the evolution of the thermal energy in the shocked wind bubble, $E_{\rm{b}}$, which evolves according to equation 31 of FGQ12: 

\begin{equation}\label{bubble_thermal_eqn} 
\dot{E}_{\rm{b}} = \frac{1}{2} \dot{M}_{\rm{in}} v_{\rm{in}}^{2} - 4 \pi R_{\rm{s}}^{2} (P_{\rm{b}} - P_{0}) \dot{R}_{\rm{s}} - L_{\rm{b}}. 
\end{equation} 
We include the ambient medium pressure, $P_{0}$, in this equation, which was missing in FGQ12, although we find that $P_{0} \ll P_{\rm{b}}$, so this change is not significant. 

The shocked wind bubble cools radiatively at a rate $L_{\rm{b}}$ due to inverse Compton cooling and free-free emission. FGQ12 highlighted that, at the high temperatures ($\ga 10^{9} \, \rm{K}$) reached by AGN shocks, the wind bubble will develop a two-temperature (2T) structure, with proton and electron temperatures $T_{\rm{p}}$ and $T_{\rm{e}}$, respectively. FGQ12 showed how to model 2T effects in spherically symmetric AGN wind models. However, we have developed a more general prescription to approximate 2T cooling, which can also be applied to 3D hydrodynamic simulations. We will adopt this more general prescription for the analytic models in this work. In Appendix~\ref{2T_cool_sect} we derive the volumetric cooling rate, $\Lambda_{\rm{IC}, \, \rm{2T}}$, due to inverse Compton cooling in a 2T plasma. Note that this 2T prescription was not used in the simulations that we ran in Paper~\textsc{i}. However, this does not affect our results, as we find in the analytic model that the outflow is still energy-driven when we use this presciption. We use a Compton temperature $T_{\rm{C}} = 2 \times 10^{7} \, \rm{K}$ \citep{sazonov04}. 

We also include free-free cooling in the shocked wind bubble, for which the cooling rate per unit volume is \citep[e.g.][]{shapiro87}: 

\begin{align}\label{free_free_eqn} 
\Lambda_{\rm{ff}}(T, n_{\rm{e}}, n_{\rm{H}}) = &1.426 \times 10^{-27} T^{1/2} n_{\rm{e}} n_{\rm{H}} \nonumber \\ 
 &\times \left[ g_{\rm{ff}} (1, T) + 0.4 g_{\rm{ff}} (2, T) \right] \, \rm{erg} \, \rm{cm}^{-3} \, \rm{s}^{-1}, 
\end{align} 
where $n_{\rm{e}}$ and $n_{\rm{H}}$ are the electron and total hydrogen number densities, respectively, and $T$ is the gas temperature. The function $g_{\rm{ff}}(Z_{i}, T)$ is given by: 

\begin{equation}\label{g_ff_eqn} 
  g_{\rm{ff}}(Z_{i}, T) = 
  \begin{cases} 
    0.79464 + 0.1243 \\ 
    \hspace{0.2in} \times \log_{10} (T / Z_{i}^{2}) & (T / Z_{i}^{2}) < 3.2 \times 10^{5} \, \rm{K} \\ 
    2.13164 - 0.1240 \\ 
    \hspace{0.2in} \times \log_{10} (T / Z_{i}^{2}) & (T / Z_{i}^{2}) \geq 3.2 \times 10^{5} \, \rm{K}, 
  \end{cases} 
\end{equation} 
where $Z_{i}$ is the ion charge of species $i$. We include only hydrogen and helium (which dominate the free-free emission, even at solar metallicity), assuming $n_{\rm{He}} / n_{\rm{H}} = 0.1$, and we have assumed in equation~\ref{free_free_eqn} that the hydrogen and helium is fully ionized. Note that, for the 2T cooling rate that we derive in Appendix~\ref{2T_cool_sect}, we assumed a pure-hydrogen plasma. However, considering the approximations that go into the 2T cooling rate, this small inconsistency is within the uncertainties for this approximate prescription. 

The total cooling rate of the shocked wind, $L_{\rm{b}}$ in equation~\ref{bubble_thermal_eqn}, can then be found by multiplying the volumetric cooling rates by the volume of the shocked wind bubble: 

\begin{equation}\label{Lb_eqn} 
L_{\rm{b}} = \frac{4 \pi}{3} R_{\rm{s}}^{3} (C \Lambda_{\rm{IC}, \, \rm{2T}}(T_{\rm{b}}, n_{\rm{b}, \, \rm{H}})\rvert_{R = R_{\rm{s}}} + \Lambda_{\rm{ff}}(T_{\rm{b}}, n_{\rm{b}, \, \rm{e}}, n_{\rm{b}, \, \rm{H}})),
\end{equation} 
where subscripts $b$ indicate quantities evaluated in the shocked wind bubble. We evaluate the 2T inverse Compton cooling rate at the outer radius of the shocked wind bubble, $R_{\rm{s}}$. However, the Compton cooling rate varies with radius, $R$. This is accounted for with the factor $C$, which depends on whether we evaluate the standard inverse Compton cooling rate using the proton temperature, $T_{\rm{p}}$, or the equilibrium electron temperature, $T_{\rm{e}}^{\rm{eq}}$ (see equation~\ref{lambda_2T_eqn}). If we assume that the proton temperature and electron density are uniform throughout the shocked wind bubble (which we found was approximately valid in the hydrodynamic simulations with the single-temperature approximation; see, for example, fig. 1 in Paper~\textsc{i}), then, using $T_{\rm{p}}$, we find that $\Lambda_{\rm{IC}, \, \rm{2T}} \propto R^{-2}$ (see equation 2.6 of Paper~\textsc{i}). However, from equation~\ref{Te eq} we see that $T_{\rm{e}}^{\rm{eq}} \propto U_{\rm{ph}}^{-2/5} \propto R^{4/5}$, where $U_{\rm{ph}}$ is the energy density of the radiation field. If $T_{\rm{e}}^{\rm{eq}} \gg T_{\rm{C}}$, then, when we use $T_{\rm{e}}^{\rm{eq}}$ in the standard Compton cooling rate, we find that $\Lambda_{\rm{IC}, \, \rm{2T}} \propto T_{\rm{e}}^{\rm{eq}} R^{-2} \propto R^{-6/5}$. By integrating $\Lambda_{\rm{IC}, \, \rm{2T}}$ over the volume of the shocked wind bubble for these two cases, we can show that $C$ is given by: 

\begin{equation} 
  C = 
  \begin{cases} 
    \frac{5}{3} & 10T_{\rm C} < T_{\rm e}^{\rm eq} \leq T_{\rm p} \\ 
    3 & \rm{otherwise}. \\ 
  \end{cases} 
\end{equation} 

Note that, unlike FGQ12, we assumed in equation~\ref{Lb_eqn} that $R_{\rm{sw}} \ll R_{\rm{s}}$. We found this necessary because calculating $R_{\rm{sw}}$ from equation 6 of FGQ12 could become numerically unstable at early times. However, we find that, apart from at early times, assuming $R_{\rm{sw}} \ll R_{\rm{s}}$ has little effect on our results. For example, at $R_{\rm{s}} > 30 \, \rm{pc}$, $v_{\rm{s}}$ differs by less than 5 per cent. 

To calculate the cooling rates, we need to calculate the temperature and density of the shocked wind bubble: 

\begin{equation}\label{sw_temp_eqn}
T_{\rm{b}} = \frac{28}{69} \frac{E_{\rm{b}} m_{\rm{p}}}{f_{\rm{mix}} M_{\rm{sw}} k_{\rm{B}}}, 
\end{equation} 
\begin{equation}\label{sw_dens_eqn} 
n_{\rm{p}} = n_{\rm{b}, \, \rm{H}} = \frac{3 f_{\rm{mix}} M_{\rm{sw}} X_{\rm{H}}}{4 \pi R_{\rm{s}}^{3} m_{\rm{p}}}, 
\end{equation} 
where $X_{\rm{H}}$ is the hydrogen mass fraction, $m_{\rm{p}}$ is the proton mass, $M_{\rm{sw}} = \dot{M}_{\rm{in}} t$ is the total mass that has been injected by the AGN after time $t$, and $f_{\rm{mix}} = (M_{\rm{sw}} + M_{\rm{cold}}) / M_{\rm{sw}}$ parameterises the mass, $M_{\rm{cold}}$, of cold gas that is mixed into the hot wind bubble, as in FGQ12. We take a fiducial value of $f_{\rm{mix}} = 1$ (i.e. no mixing). FGQ12 showed that, for large $v_{\rm{in}}$, the choice of $f_{\rm{mix}}$ is unimportant so long as it is not very large, because a lot of mixing is necessary to cause the hot wind bubble to cool. For a fully ionized plasma with $n_{\rm{He}} / n_{\rm{H}} = 0.1$, the electron density, as used for the free-free cooling rate, is then $n_{\rm{e}} = 1.2 n_{\rm{p}}$. The mean molecular weight is $\mu = 14/23$, which leads to the numerical pre-factor in equation~\ref{sw_temp_eqn}. 

The pressure of the shocked wind bubble can then be calculated from $E_{\rm{b}}$ as: 

\begin{equation}\label{bubble_pressure_eqn} 
P_{\rm{b}} = \frac{E_{\rm{b}}}{2 \pi R_{\rm{s}}^{3}}, 
\end{equation} 
where we again assume that $R_{\rm{sw}} \ll R_{\rm{s}}$. 

The FGQ12 model accounts for radiative cooling in the shocked wind bubble, but it does not include cooling in the shocked ISM layer. However, in this work we want to use this model to predict when the shocked ISM layer will cool. We therefore extended the FGQ12 model to track the thermal and kinetic energies of the shocked ISM layer, $E_{\rm{s}, \, \rm{th}}$ and $E_{\rm{s}, \, \rm{kin}}$ respectively. 

The total energy in this layer, $E_{\rm{s}, \, \rm{tot}} = E_{\rm{s}, \, \rm{th}} + E_{\rm{s}, \, \rm{kin}}$, evolves according to: 

\begin{equation}~\label{cool_s_eqn}  
\dot{E}_{\rm{s}, \, \rm{tot}} = 4 \pi R_{\rm{s}}^{2} (P_{\rm{b}} - P_{0}) \dot{R}_{\rm{s}} - \frac{G M_{\rm{s}} M_{\rm{t}}}{R_{\rm{s}}^{2}} \dot{R}_{\rm{s}} - L_{\rm{s}}. 
\end{equation} 
The first two terms on the right hand side of equation~\ref{cool_s_eqn} account for the work done on the shocked ISM layer by the shocked wind bubble pressure and the gravitational potential. 

To model the radiative cooling rate in this layer, $L_{\rm{s}}$, we include two cooling processes. Firstly, we include free-free emission, using equation~\ref{free_free_eqn} for the volumetric free-free cooling rate. The free-free cooling rate of the whole shocked ISM layer can then be found by multiplying equation~\ref{free_free_eqn} by the volume of this layer: 

\begin{align} 
L_{\rm{s}, \, \rm{ff}} &= \Lambda_{\rm{ff}}(T_{\rm{s}}, n_{\rm{s}, \rm{e}}, n_{\rm{s}, \, \rm{H}}) \frac{M_{\rm{s}} X_{\rm{H}}}{m_{\rm{p}} n_{\rm{s}, \, \rm{H}}} \nonumber \\
 &= \frac{1.426 \times 10^{-27} T_{\rm{s}}^{1/2} n_{\rm{s}, \, \rm{e}} M_{\rm{s}} X_{\rm{H}}}{m_{\rm{p}}} \nonumber \\ 
 & \hspace{0.7in} \times \left[ g_{\rm{ff}} (1, T_{\rm{s}}) + 0.4 g_{\rm{ff}} (4, T_{\rm{s}}) \right] \, \rm{erg} \, \rm{s}^{-1}, 
\end{align} 
where $T_{\rm{s}}$, $n_{\rm{s}, \, \rm{e}}$ and $n_{\rm{s}, \, \rm{H}}$ are the temperature, electron density and hydrogen density, respectively, of the shocked ISM layer. 

We also include metal line cooling, which dominates below $\sim 10^{7} \, \rm{K}$. This can be approximated by a piecewise power law fit \citep[e.g.][]{maclow88, draine11}: 

\begin{equation} \label{lambda_line_eqn} 
  \Lambda_{\rm{line}} = 
  \begin{cases} 
    0 &T_{\rm{s}} \leq 10^{4} \, \rm{K} \\
    5.0 \times 10^{-22} \left( \frac{T_{\rm{s}}}{10^{5} \, \rm{K}} \right)^{2} \left( \frac{Z}{\rm{Z}_{\odot}} \right) \\ 
    \hspace{0.4in} \times n_{\rm{s}, \, \rm{e}} n_{\rm{s}, \, \rm{H}} \, \rm{erg} \, \rm{cm}^{-3} \, \rm{s}^{-1} & \hspace{-0.07in} 10^{4} < T_{\rm{s}} \leq 10^{5} \, \rm{K} \\
    5.0 \times 10^{-22} \left( \frac{T_{\rm{s}}}{10^{5} \, \rm{K}} \right)^{-0.7} \left( \frac{Z}{\rm{Z}_{\odot}} \right) \\ 
    \hspace{0.4in} \times n_{\rm{s}, \, \rm{e}} n_{\rm{s}, \, \rm{H}} \, \rm{erg} \, \rm{cm}^{-3} \, \rm{s}^{-1} &T_{\rm{s}} > 10^{5} \, \rm{K}. 
  \end{cases} 
\end{equation} 
For $T_{\rm{s}} > 10^{5} \, \rm{K}$, we take this from equation 34.2 of \citet{draine11}, which they fit to the cooling function in their fig. 34.1, calculated at solar metallicity assuming collisional ionization equilibrium (CIE). We then approximately fit a second power law to this cooling function in the range $10^{4} < T_{\rm{s}} \leq 10^{5} \, \rm{K}$. 

Following \citet{maclow88}, we scale the metal line cooling rate linearly with metallicity, $Z$. This assumption will break down at low metallicity, when line cooling from hydrogen and helium dominate. However, it is sufficient for metallicities $> 0.1 \, \rm{Z}_{\odot}$ that we consider in this work. \citet{draine11} use solar abundances from \citet{asplund09}, with $\rm{Z}_{\odot, \, \rm{Asplund}} = 0.0142$. We have therefore renormalised their equation to our adopted solar metallicity of $\rm{Z}_{\odot} = 0.0129$, from table 1 of \citet{wiersma09}. 

Photoionization and non-equilibrium ionization will affect the metal line cooling \citep{efstathiou92, gnat07, wiersma09, oppenheimer13, richings14a}. However, we show in section~\ref{comparison_sect} that this approximate cooling function in CIE is sufficient to reproduce the behaviour of the simulations. This is likely because photoionization only becomes important at $T \la 10^{7} \, \rm{K}$, where metal line cooling is significant. Above this temperature, cooling is dominated by free-free emission, which is not strongly affected by photoionization. Since it is the onset of significant cooling that determines when the cool gas forms, we expect CIE to be a good approximation for shock temperatures $T_{\rm{sh}}(v_{\rm{sh}}) \ga 10^{7} \, \rm{K}$, i.e. for shock velocities $v_{\rm{sh}} \ga 1000 \, \rm{km} \, \rm{s}^{-1}$. 

The metal line cooling rate of the shocked ISM layer is then: 

\begin{equation}\label{line_cooling_eqn}  
  L_{\rm{s}, \, \rm{line}} = 
  \begin{cases} 
    0 &T_{\rm{s}} \leq 10^{4} \, \rm{K} \\ 
    5.0 \times 10^{-22} \left( \frac{T_{\rm{s}}}{10^{5} \, \rm{K}} \right)^{\beta} \left( \frac{Z}{\rm{Z}_{\odot}} \right) \\ 
    \hspace{0.4in} \times \frac{n_{\rm{s}, \, \rm{e}} M_{\rm{s}} X_{\rm{H}}}{m_{\rm{p}}} \, \rm{erg} \, \rm{s}^{-1} &T_{\rm{s}} > 10^{4} \, \rm{K}, 
  \end{cases} 
\end{equation} 
where: 

\begin{equation} 
  \beta = 
  \begin{cases} 
    2  &10^{4} < T_{\rm{s}} \leq 10^{5} \, \rm{K} \\ 
    -0.7  &T_{\rm{s}} > 10^{5} \, \rm{K}. 
  \end{cases} 
\end{equation} 

The total cooling rate is then: 

\begin{equation} 
L_{\rm{s}} = L_{\rm{s}, \, \rm{ff}} + L_{\rm{s}, \, \rm{line}}.
\end{equation} 

We have truncated the cooling function (both from free-free and metal line cooling) at $10^{4} \, \rm{K}$. Below this temperature, molecular cooling becomes important. For example, we saw in Paper \textsc{i} that there is strong H$_{2}$ emission from molecular gas at temperatures above a few hundred K. However, we do not follow molecule formation in this simple model. Instead, we are interested in whether the shocked ISM layer is able to cool down from the post-shock temperature ($\sim 10^{7-8} \, \rm{K}$) to $\sim 10^{4} \, \rm{K}$. Once the layer has cooled to $~10^{4} \, \rm{K}$, we use the hydro-chemical simulations from Paper~\textsc{i} to estimate the molecular gas fraction in the swept up gas (see Section~\ref{obs_sect}).

To calculate $L_{\rm{s}}$, we need to know the temperature and density of the shocked ISM layer. The temperature can be calculated from the thermal energy: 

\begin{equation}\label{Ts_eqn} 
T_{\rm{s}} = \frac{28}{69} \frac{E_{\rm{s}, \, \rm{th}} m_{\rm{p}}}{M_{\rm{s}} k_{\rm{B}}}, 
\end{equation} 
where we again assume that the gas is a fully ionized hydrogen plus helium plasma, with $\mu = 14 / 23$. 

We assume that this layer remains in pressure equilibrium with the shocked wind bubble, with $P_{\rm{s}} = P_{\rm{b}}$. In other words, as this layer cools, its density increases, and its radial thickness decreases. This behaviour is consistent with the simulations in Paper \textsc{i}, although the temperature-density diagrams show a large scatter around pressure equilibrium in this layer (see fig. 5 in Paper \textsc{i}). The hydrogen number density is then: 

\begin{equation} 
n_{\rm{s}, \, \rm{H}} = \frac{3}{2} \frac{P_{\rm{b}} M_{\rm{s}} X_{\rm{H}}}{m_{\rm{p}} E_{\rm{s}, \, \rm{th}}}, 
\end{equation} 
and the electron number density is $n_{\rm{s}, \, \rm{e}} = 1.2 n_{\rm{s}, \, \rm{H}}$. 

The kinetic energy in outward bulk motion of the shocked ISM layer is: 

\begin{equation}\label{kin_eqn} 
E_{\rm{s}, \, \rm{kin}} = \frac{1}{2} M_{\rm{s}} v_{\rm{s}}^{2}. 
\end{equation} 
Differentiating equation~\ref{kin_eqn} with respect to time, we get: 

\begin{align}\label{kin_dot_eqn} 
\dot{E}_{\rm{s}, \, \rm{kin}} &= M_{\rm{s}} v_{\rm{s}} \dot{v}_{\rm{s}} + \frac{1}{2} \dot{M}_{\rm{s}} v_{\rm{s}}^{2} \nonumber \\ 
 &= M_{\rm{s}} v_{\rm{s}} \dot{v}_{\rm{s}} + 2 \pi R_{\rm{s}}^{2} \rho_{\rm{g}} (R_{\rm{s}}) v_{\rm{s}}^{3}. 
\end{align} 
Then the thermal energy of the shocked ISM layer evolves according to: 

\begin{equation}\label{th_dot_eqn} 
\dot{E}_{\rm{s}, \, \rm{th}} = \dot{E}_{\rm{s}, \, \rm{tot}} - \dot{E}_{\rm{s}, \, \rm{kin}}. 
\end{equation} 
We therefore evolve $E_{\rm{s}, \, \rm{th}}$ in time using equations~\ref{cool_s_eqn}, \ref{kin_dot_eqn} and \ref{th_dot_eqn}, and then calculate $E_{\rm{s}, \, \rm{kin}}$ using equation~\ref{kin_eqn}.

For each model, we specify $L_{\rm{AGN}}$, $\rho_{0} (R_{0} = 1 \, \rm{kpc})$ (which we give in terms of the ambient hydrogen density, $n_{\rm{H}0}$), $\alpha$ and $Z$. Equations~\ref{eqn_of_motion}, \ref{bubble_pressure_eqn} and \ref{cool_s_eqn} are not valid at $R_{\rm{s}} = 0$, so we start at a small, finite radius $R_{\rm{s}} = 0.1 \, \rm{pc}$, with $v_{\rm{s}} = v_{\rm{in}}$ and $E_{\rm{b}} = E_{\rm{s}, \, \rm{th}} = 0$. We then integrate equations ~\ref{eqn_of_motion}, \ref{bubble_thermal_eqn} and \ref{th_dot_eqn} in time. We find it necessary to limit $v_{\rm{s}} \leq v_{\rm{in}}$, otherwise $v_{\rm{s}}$ can increase by an order of magnitude at early times. However, capping $v_{\rm{s}}$ in this way has little effect on the late-time evolution. For example, at $R_{\rm{s}} > 4 \, \rm{pc}$, $v_{\rm{s}}$ changes by less than 1 per cent if it has been capped at early times. 

\section{Comparison of simulations with the analytic model}\label{comparison_sect} 

In this section we compare our simulations from Paper~\textsc{i} to the analytic model described in Section~\ref{analytic_model_sect}. We ran the analytic model four times, with the same parameters as the simulations, and with a density slope $\alpha = 0$, i.e. for a uniform ambient ISM. We use the simulations at the standard resolution level ($30 \, \rm{M}_{\odot}$ per gas particle, with a minimum gravitational softening length for gas particles of $0.1 \, \rm{pc}$) from Paper~\textsc{i}. We show in Appendix~\ref{resolution_appendix} that the results presented in this section are well converged with numerical resolution.  

\begin{figure}
\centering
\mbox{
	\includegraphics[width=84mm]{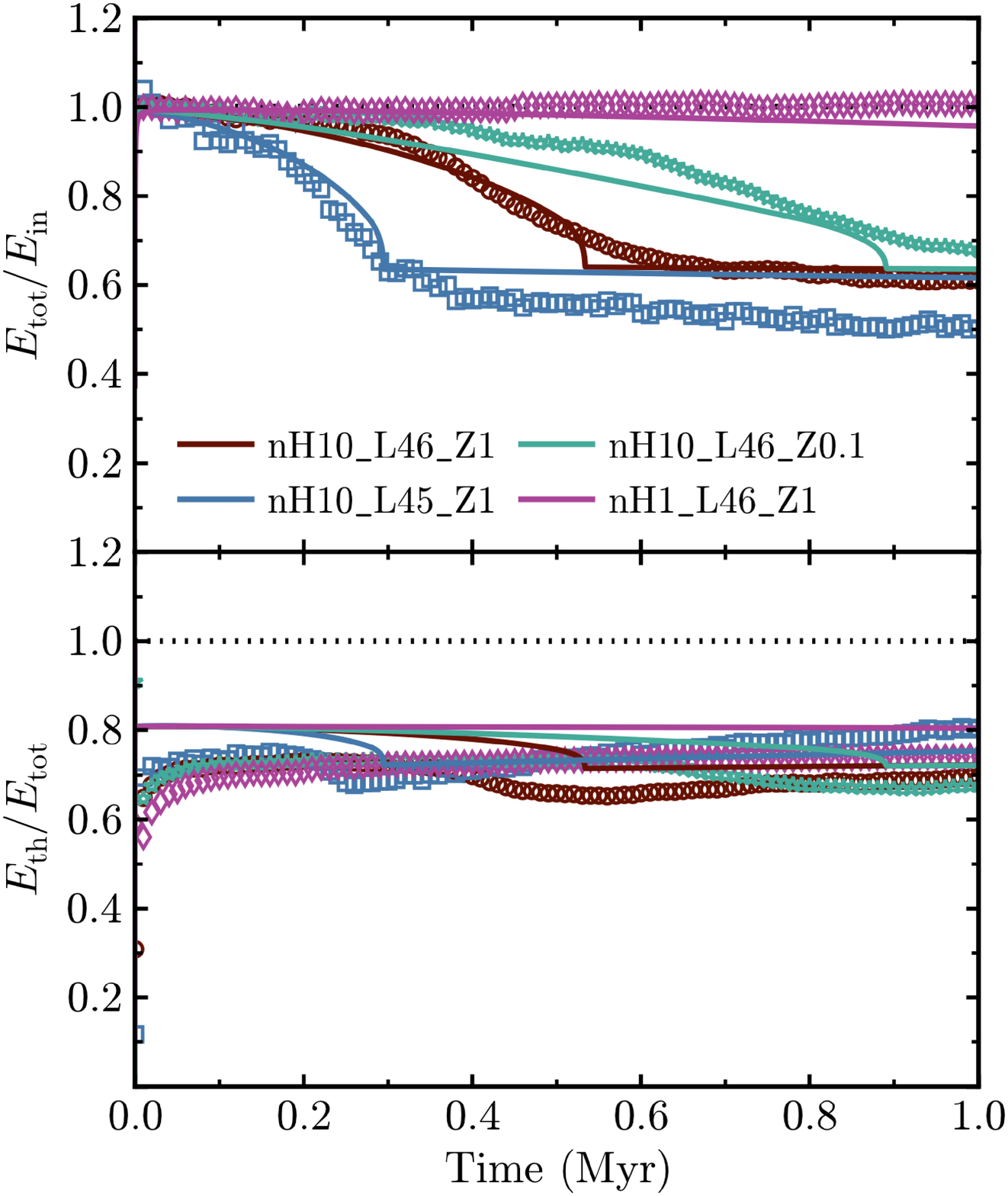}}
\caption{The ratio of total outflow energy to mechanical wind energy injected by the AGN ($E_{\rm{tot}} / E_{\rm{in}}$; top panel), and the ratio of thermal to total outflow energy ($E_{\rm{th}} / E_{\rm{tot}}$; bottom panel), versus time. We show the analytic model (solid curves) and the simulations (symbols) for runs nH10\_L46\_Z1 (red), nH10\_L45\_Z1 (blue), nH10\_L46\_Z0.1 (green), and nH1\_L46\_Z1 (magenta). The analytic model reproduces the energy losses in the simulations, which are primarily due to radiative cooling in the shocked ISM layer.} 
\label{energySimFig}
\end{figure}

The top panel of Fig.~\ref{energySimFig} shows the time evolution of the total (thermal plus kinetic, $E_{\rm{tot}}$) energy of the outflow, in both the hot wind bubble and the shocked ISM layer, normalised by the integrated mechanical wind energy injected by the AGN after time $t$, i.e. $E_{\rm{in}} = \frac{1}{2} \dot{M}_{\rm{in}} v_{\rm{in}}^{2} t$. The solid curves show the analytic models, while the symbols show the simulations. In the simulations, we calculate the total energy in the full simulation box, rather than only the high-resolution region. In Paper~\textsc{i}, we limited our analysis to a high-resolution wedge, i.e. particles within the high-resolution octant with polar and azimuthal angles in spherical polar coordinates between $15^{\circ}$ and $75^{\circ}$. This was necessary to avoid artifacts along the boundaries between the high- and low-resolution regions. However, we found that, if we calculate the energy of the outflow only in the high-resolution wedge and then scale up to the full box by multiplying by the ratio of the solid angle of a sphere to the solid angle subtended by the wedge, then the total energy exceeds the injected energy by up to 40 per cent. This suggests that there is a net energy flux from the low- to the high-resolution region. It is not clear what the cause of this energy flux is, although it may be related to the artifacts along the boundaries between these two regions. Therefore, throughout this section we use the full simulation box for our analysis. We found that using the full box, rather than the high-resolution wedge, does not strongly affect the rest of the results in this section. 

Initially, the outflow is energy-conserving ($E_{\rm{tot}} / E_{\rm{in}} = 1$). Once the shocked ISM layer is able to cool, $E_{\rm{tot}} / E_{\rm{in}}$ decreases. The low-luminosity run (blue curve/symbols) cools fastest, while the low-density run (magenta) does not cool after $1 \, \rm{Myr}$. There is good agreement between the simulations and the analytic model for the time at which each run starts to cool. 

Once the shocked ISM layer has radiated away most of its thermal energy, the evolution of $E_{\rm{tot}} / E_{\rm{in}}$ flattens out at a value that corresponds to the total kinetic energy (of both the hot wind bubble and the shocked ISM layer) plus the thermal energy of the hot wind bubble, which does not cool in any of these runs within $1 \, \rm{Myr}$. In the three analytic model runs for which the shocked ISM layer cools within $1 \, \rm{Myr}$, the total energy flattens out at $E_{\rm{tot}} / E_{\rm{in}} \approx 0.6$. This is in good agreement with the fiducial simulation (red), although the low-luminosity simulation (blue) flattens out at $E_{\rm{tot}} / E_{\rm{in}} = 0.5$. In the low-metallicity simulation (green), $E_{\rm{tot}} / E_{\rm{in}}$ is still declining after $1 \, \rm{Myr}$. Because the pressure of the hot shocked wind bubble determines the overall dynamical evolution of the outflow, all these models are considered energy-conserving even though the outer, shocked ISM layer does cool.

The energy losses in the top panel of Fig.~\ref{energySimFig} could also be caused by work done against the gravitational potential. However, in the analytic model, we find that 90 per cent of the energy losses in the fiducial and low-metallicity runs, and 80 per cent in the low-luminosity run, are due to radiative cooling in the shocked ISM layer. 

The bottom panel of Fig.~\ref{energySimFig} shows the ratio of the thermal ($E_{\rm{th}}$) to total energy of the outflow versus time. In the analytic models, initially $E_{\rm{th}} / E_{\rm{tot}} \approx 0.8$. This decreases to $\approx 0.7 - 0.75$ after $1 \, \rm{Myr}$, except for the low-density run, which does not cool within this time. At early times, the fraction of thermal energy in the simulations is lower than the analytic models. However, at times $\ga 0.05 \, \rm{Myr}$, the simulations have $E_{\rm{th}} / E_{\rm{tot}} \approx 0.65 - 0.8$, in reasonable agreement with the analytic models. 

\begin{figure}
\centering
\mbox{
	\includegraphics[width=84mm]{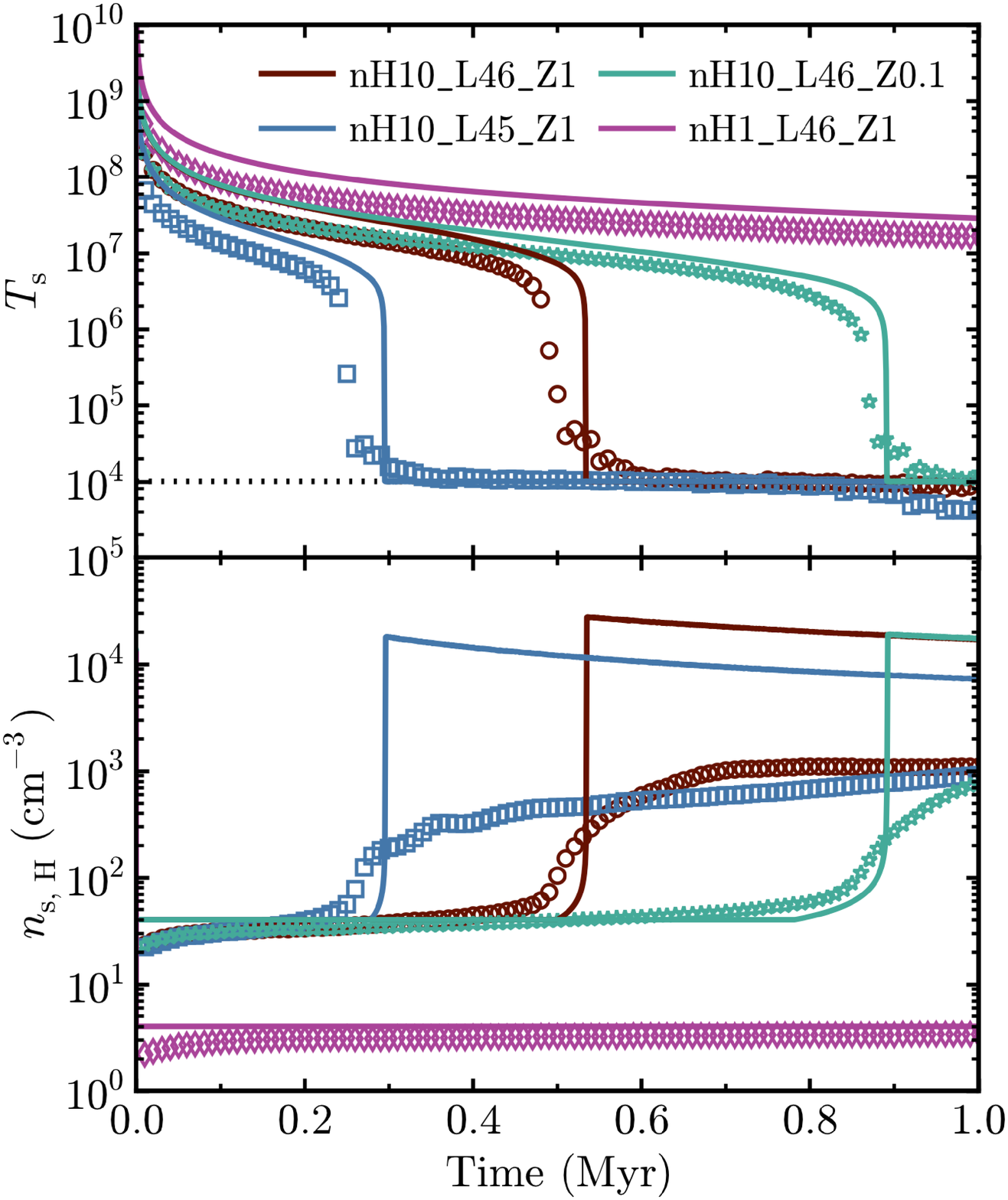}}
\caption{Time evolution of the temperature ($T_{\rm{s}}$; top panel) and hydrogen density ($n_{\rm{s}, \, \rm{H}}$; bottom panel) of the shocked ISM layer in the analytic model (solid curves) and the simulations (symbols). In the simulations, $T_{\rm{s}}$ and $n_{\rm{s}, \, \rm{H}}$ are the median temperature and hydrogen density, respectively, of particles with densities $> 2 n_{\rm{H}0}$. This density cut corresponds to the shocked ISM layer (see text). The analytic model correctly predicts the cooling time of the shocked ISM layer from the simulations, although it overpredicts the density after this layer has cooled, due to the assumption of pressure equilibrium.} 
\label{TSimFig}
\end{figure}

In Fig.~\ref{TSimFig} we show how the temperature ($T_{\rm{s}}$; top panel) and the hydrogen density ($n_{\rm{s}, \, \rm{H}}$; bottom panel) of the shocked ISM layer evolves with time in the analytic models (solid curves) and the simulations (symbols). In the simulations, we define particles with densities $> 2 n_{\rm{H}0}$ to be in the shocked ISM layer. Particles in the ambient ISM that have not yet been swept up by the outflow are very close to $n_{\rm{H}0}$, while particles in the hot wind bubble have much lower densities (see fig.~1 in Paper~\textsc{i}). We then take $T_{\rm{s}}$ and $n_{\rm{s}, \, \rm{H}}$ to be the median temperature and hydrogen density, respectively, of particles in the shocked ISM layer. 

The horizontal dotted line in the top panel of Fig.~\ref{TSimFig} indicates a temperature of $10^{4} \, \rm{K}$, below which we truncate the radiative cooling function. Note that $T_{\rm{s}}$ can still fall below $10^{4} \, \rm{K}$. For example, we see from equation~\ref{Ts_eqn} that, if $M_{\rm{s}}$ rises more rapidly than $E_{\rm{s}, \, \rm{th}}$, then $T_{\rm{s}}$ will decrease even without radiative cooling. 

Initially, the swept up gas is shocked-heated to $\sim 10^{8} - 10^{9} \, \rm{K}$. $T_{\rm{s}}$ then gradually declines, because, as noted above, $T_{\rm{s}}$ is also sensitive to how $M_{\rm{s}}$ and $E_{\rm{s}, \, \rm{th}}$ evolve, and so $T_{\rm{s}}$ can still evolve even in the absence of radiative losses. Physically, this decline in $T_{\rm{s}}$ reflects the facts that, as the outflow decelerates, the post-shock temperature of the swept up gas also decreases. Once $T_{\rm{s}}$ reaches $\sim 10^{6.5} \, \rm{K}$, radiative cooling becomes efficient and $T_{\rm{s}}$ drops rapidly to $10^{4} \, \rm{K}$. 

The analytic model generally predicts a temperature $T_{\rm{s}}$ that is slightly higher (typically by $\la 0.5 \, \rm{dex}$) than in the simulations. However, the time at which $T_{\rm{s}}$ falls to $10^{4} \, \rm{K}$ in the analytic model is in good agreement with the simulations. We found in the simulations that, once the shocked ISM layer has cooled below $10^{4} \, \rm{K}$, H$_{2}$ can then form rapidly. This confirms a common assumption used in previous theoretical works that predicted the formation of molecular gas in AGN-driven outflows but did not actually follow the time-dependent chemistry \citep[e.g.][]{zubovas14}. We can therefore use the analytic model to predict when an AGN wind is likely to form a molecular outflow (assuming that dust grains are present in the outflow, as assumed in our simulations). 

Before cooling becomes significant, the shocked ISM density in the analytic model is $4 n_{\rm{H}0}$, as expected for strong shocks. This is in good agreement with the simulations, except that in the simulations it takes $\sim 0.1 \, \rm{Myr}$ for the density to build up. However, after the shocked ISM layer has cooled to $10^{4} \, \rm{K}$, the analytic model overpredicts the density by more than an order of magnitude. This discrepancy arises because the analytic model assumes that the shocked ISM layer is in pressure equilibrium with the hot wind bubble. But in the temperature-density diagrams from the simulations (fig.~5 in Paper~\textsc{i}), we see that the shocked ISM pressure is less than the pressure of the hot wind bubble. 

\begin{figure}
\centering
\mbox{
	\includegraphics[width=84mm]{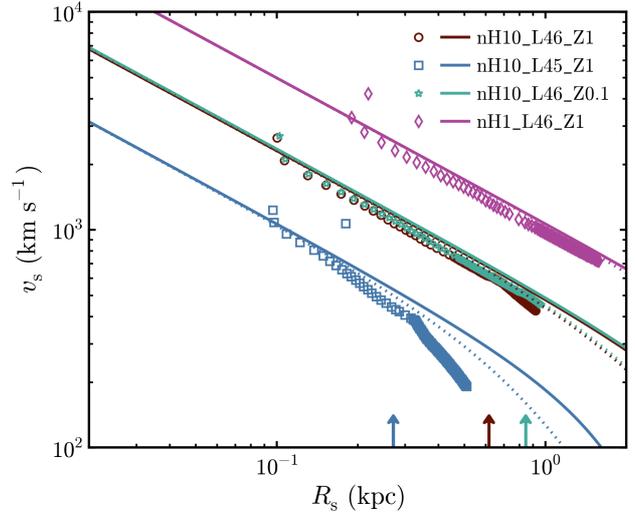}}
\caption{Evolution of the forward shock velocity, $v_{\rm{s}}$, with radius, $R_{\rm{s}}$, in the analytic model (solid curves) and the simulations (symbols). The dotted curves show the analytic model when we account for the inward momentum of the ambient medium under the influence of gravity. The arrows show the radius at which the corresponding analytic model first cools to $10^{4} \, \rm{K}$. In the simulations, $v_{\rm{s}}$ and $R_{\rm{s}}$ in a given snapshot are the mass-weighted mean radial velocity and radius, respectively, of particles in that snapshot with densities $> 2 n_{\rm{H}0}$. The analytic model (solid curves) is in good agreement with the simulations, except in the low-luminosity run, for which $v_{\rm{s}}$ in the simulations falls off more steeply with $R_{\rm{s}}$ than in the analytic model. This discrepancy is at least partly due to the effects of gravity on the ambient medium (see text).} 
\label{RvSimFig}
\end{figure}

In Fig.~\ref{RvSimFig} we show how the velocity of the forward shock, $v_{\rm{s}}$, evolves with radius, $R_{\rm{s}}$. We calculate $v_{\rm{s}}$ and $R_{\rm{s}}$ in the simulations to be the mass-weighted mean radial velocity and radius, respectively, of particles in a given snapshot with densities $> 2 n_{\rm{H}0}$, which corresponds to the shocked ISM layer. The arrows show the radius at which the corresponding analytic model first cools to $10^{4} \, \rm{K}$. There is generally good agreement between the analytic model (solid curves) and the simulations (symbols), except for the low-luminosity run (blue), where the velocity in the simulations falls off more steeply with radius than the analytic model at $R_{\rm{s}} \ga 0.2 \, \rm{kpc}$. This is at least partly explained by the effects of gravity on the ambient medium in the simulations. In the analytic model, the ambient medium remains stationary until it is swept up by the outflow. However, while the ambient medium in the simulations is initially stationary, it subsequently moves inwards due to the gravitational potential of the black hole and the host galaxy. The dotted curves show the analytic model when we account for the inward momentum of the swept up gas due to gravity. This tends to slow down the outflow, and is more noticeable for the low-luminosity run because the lower outflow velocities in this run are more susceptible to this effect. However, we note that this effect is an unphysical consequence of the idealised setup of our simulations. In a realistic galaxy, the ambient medium would be supported, for example, by rotation, and so would not form a strong inflow. 

For a given $R_{\rm{s}}$, $v_{\rm{s}}$ increases with decreasing $n_{\rm{H}0}$ (in agreement with FGQ12) and increasing $L_{\rm{AGN}}$, and is independent of metallicity. 

\begin{figure}
\centering
\mbox{
	\includegraphics[width=84mm]{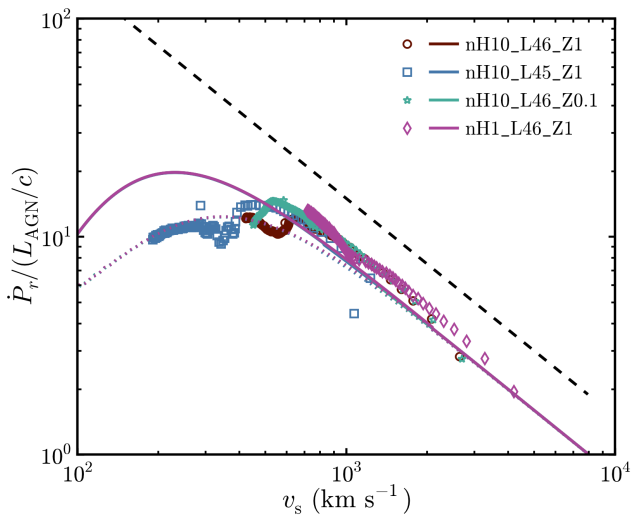}}
\caption{Instantaneous rate of change of radial momentum ($\dot{P}_{r}$) normalised by the AGN momentum injection rate ($L_{\rm{AGN}} / c$), i.e. the momentum boost factor, plotted against the forward shock velocity ($v_{\rm{s}}$), for the analytic models (solid curves) and the simulations (symbols). The dashed black line shows the expected relation for an energy-conserving outflow, assuming that half the outflow energy is kinetic. At high $v_{\rm{s}}$ (early times), the analytic models and simulations follow the slope of the energy-conserving relation, and are in good agreement with one another. The momentum boost peaks at $\approx 10$ (simulations) or $\approx 20$ (analytic models). When we account for the effects of gravity on the ambient medium in the analytic model (dotted curves), the momentum boost peaks at $\approx 10$, in agreement with the simulations.} 
\label{PrDotSimFig}
\end{figure}

If the outflow is energy-conserving, the thermal pressure of the hot wind bubble will accelerate the outflow, boosting its momentum beyond that of the small-scale AGN wind (FGQ12; \citealt{zubovas12}; \citealt{costa14}). Fig.~\ref{PrDotSimFig} shows this momentum boost, defined here as the instantaneous rate of change of radial momentum of the outflow ($\dot{P}_{r}$) normalised by the momentum injection rate of the AGN ($L_{\rm{AGN}} / c$, with $\tau_{\rm{in}} = 1$), plotted against $v_{\rm{s}}$. Note that some observational papers refer to `instantaneous' rates to mean rates averaged over the time taken for the outflow to cross the thickness of the outflowing shell (see, for example, the discussion in \citealt{veilleux17}, and references therein). However, we use `instantaneous' to mean averaged over a single time-step in the integration of the analytic model. 

The solid curves and symbols in Fig.~\ref{PrDotSimFig} show the analytic model and the simulations, respectively. In the simulations we include only particles that are outflowing ($P_{r} > 0$), including the hot shocked wind bubble as well as the swept up gas, although the latter component dominates the mass and momentum of the outflow. The dotted curves show the analytic model when we include the inward momentum of the ambient medium under the influence of the gravitational potential of the black hole and the host galaxy. The dashed black line shows the expected relation for an energy-conserving outflow, assuming that half of the energy injected by the AGN wind goes into the kinetic energy of the shocked ISM layer (see equation~38 of FGQ12). 

At high $v_{\rm{s}}$ (i.e. early times), the analytic models and the simulations are in good agreement, and follow the same slope as expected for an energy-conserving outflow. We saw in Fig.~\ref{energySimFig} that, in both the simulations and the analytic models, $\approx 20 - 30$ per cent of the outflow energy is kinetic. This explains why, in Fig.~\ref{PrDotSimFig}, they are lower than the black dashed line, which assumes that 50 per cent of the energy is kinetic. 

The momentum boost in the simulations peaks at $\approx 10$, at $v_{\rm{s}} \approx 400 - 700 \, \rm{km} \, \rm{s}^{-1}$. In the analytic models (solid curves), the momentum boost continues to increase for longer, peaking at $\approx 20$, at $v_{\rm{s}} \approx 200 \, \rm{km} \, \rm{s}^{-1}$. This discrepancy between the analytic models and the simulations is due to the effects of gravity on the ambient medium, which causes the ambient medium to move inwards in the simulations, while this effect is not included by default in the analytic model. The dotted curves show that, if we do account for the inward momentum of the ambient medium under the influence of gravity in the analytic model, the momentum boost factor peaks at $\approx 10$, as seen in the simulations. However, we again stress that this effect is an unphysical consequence of the idealised setup of our simulations. We therefore do not include this effect in the analytic model for the remainder of the paper. 

In Section~\ref{param_sect}, we will show that the deviation of the momentum boost factor from the simple $\dot{P}_{r} \propto 1 / v_{\rm{s}}$ energy-conserving scaling (which continues to increase towards lower $v_{\rm{s}}$ and does not reach a peak) is due to work done by the outflow against the gravitational potential of the host galaxy and the black hole. We also see in Fig.~\ref{PrDotSimFig} that all four analytic model runs follow exactly the same relation between the momentum boost and $v_{\rm{s}}$. We will show in Section~\ref{param_sect} that, in the analytic model, this relation depends only on the density profile slope and the gravitational potential of the host galaxy and the black hole. 

\section{Parameter exploration}\label{param_sect} 

We demonstrated in the previous section that the analytic model reproduces the behaviour of the simulations. In particular, it correctly predicts the time at which the shocked ISM layer cools below $10^{4} \, \rm{K}$, as determined by the simulations (top panel of Fig.~\ref{TSimFig}). We can now use the analytic model to explore a much wider range of the parameters of the physical setup, to investigate under what conditions the shocked ISM layer can cool. 

\begin{figure}
\centering
\mbox{
	\includegraphics[width=84mm]{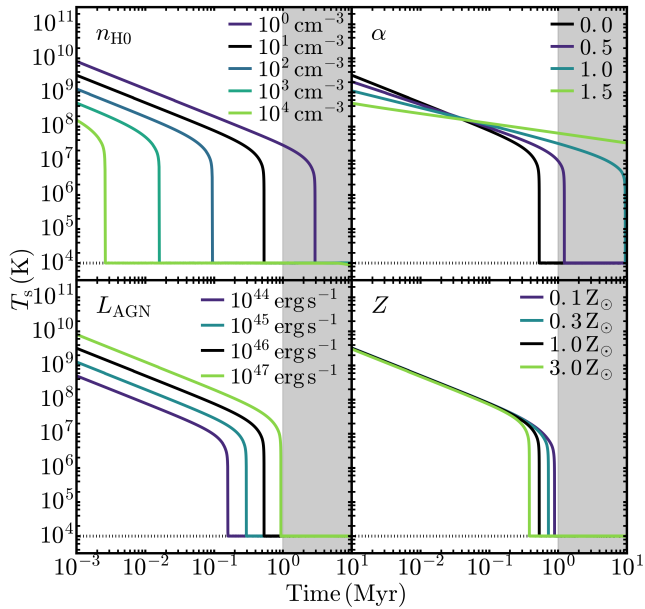}}
\caption{Temperature of the shocked ISM layer, $T_{\rm{s}}$, plotted against time, for the analytic model with variations of the ambient density at $100$ pc ($n_{\rm{H}0}$; top left), the slope of the density profile ($\alpha$; top right), the AGN luminosity ($L_{\rm{AGN}}$; bottom left), and the metallicity ($Z$; bottom right). In each panel, the parameters that are not varied are held fixed at their fiducial values: $n_{\rm{H}0} = 10 \, \rm{cm}^{-3}$, $\alpha = 0$, $L_{\rm{AGN}} = 10^{46} \, \rm{erg} \, \rm{s}^{-1}$, and $Z = \rm{Z}_{\odot}$. The grey shaded region highlights times $> 1 \, \rm{Myr}$. To reproduce observed molecular outflows, the shocked ISM layer needs to cool before this time, which corresponds to the typical flow times ($r / v$) of outflows observed in luminous quasars. The cooling time to reach $10^{4} \, \rm{K}$ decreases with increasing $n_{\rm{H}0}$ and $Z$, and with decreasing $\alpha$ and $L_{\rm{AGN}}$.} 
\label{TFig}
\end{figure} 

We ran the analytic model with the same parameters as the fiducial simulation run, i.e. $n_{\rm{H}0} = 10 \, \rm{cm}^{-3}$, $\alpha = 0$, $L_{\rm{AGN}} = 10^{46} \, \rm{erg} \, \rm{s}^{-1}$, and $Z = \rm{Z}_{\odot}$. We then varied each of these four parameters in turn, while keeping the remaining parameters fixed at their fiducial values. In Fig.~\ref{TFig} we show how $T_{\rm{s}}$ evolves with time in the analytic model, for variations of $n_{\rm{H}0}$ (top left), $\alpha$ (top right), $L_{\rm{AGN}}$ (bottom left), and $Z$ (bottom right). The grey shaded region in each panel highlights times longer than $1 \, \rm{Myr}$. Since observed molecular outflows in luminous quasars have flow times $r / v \sim 1 \, \rm{Myr}$ \citep[e.g.][]{gonzalezalfonso17}, the outflow models will need to cool within $1 \, \rm{Myr}$ to have a chance of reproducing observed molecular outflows. We note that if molecular outflows are observed on different spatial scales $r$ or time scales $r / v$ (as may be the case in different AGN samples), then the criteria for determining whether the shocked ISM layer cools fast enough to produce molecules should be modified accordingly. 

At early times, before the shocked ISM layer can radiatively cool, $T_{\rm{s}}$ declines steadily in all runs. This behaviour was also seen in the simulations (Fig.~\ref{TSimFig}), and is due to the deceleration of $v_{\rm{s}}$ as the outflow expands, which reduces the post-shock temperature of the swept up gas. Once $T_{\rm{s}}$ reaches $\sim 10^{6.5 - 7} \, \rm{K}$, radiative cooling becomes efficient and $T_{\rm{s}}$ drops rapidly to $10^{4} \, \rm{K}$, where we truncate the radiative cooling function. 

As the density, $n_{\rm{H}0}$, increases (top left panel), the time for $T_{\rm{s}}$ to cool to $10^{4} \, \rm{K}$ decreases. This trend arises for two reasons. Firstly, as the density increases, the outflow decelerates more rapidly (see Fig.~\ref{RvSimFig}, and fig.~4 of FGQ12). The lower $v_{\rm{s}}$ results in a lower post-shock temperature of the swept up gas, and so it more quickly reaches a temperature of $\sim 10^{6.5 - 7} \, \rm{K}$ where radiative cooling becomes efficient. This is also why the curves decrease in normalisation with increasing density. Secondly, as the density increases, the radiative cooling time-scale in the shocked ISM layer decreases. This allows radiative cooling to become efficient at a (slightly) higher temperature, and hence earlier. 

In the top right panel of Fig.~\ref{TFig}, the shocked ISM layer cools more slowly with increasing $\alpha$ (i.e. steeper density profiles). This is because we normalise the density profiles to $n_{\rm{H}0} = 10 \, \rm{cm}^{-3}$ at $100$ pc. In the runs where we vary $\alpha$, the shocked ISM layer cools beyond this radius, so, as we increase $\alpha$, we decrease the density at the cooling radius, which has a similar effect to decreasing $n_{\rm{H}0}$. However, we note that, if we normalised the density profile at a larger radius, beyond the cooling radius, then increasing $\alpha$ would increase the density at the cooling radius. In this case, we find the opposite trend, where increasing $\alpha$ decreases the time it takes for the shocked ISM layer to cool (not shown). We therefore caution that the trends with $\alpha$ depend on how we normalise the density profile. 

As the AGN luminosity increases (bottom left panel), the cooling time increases. This is because the AGN wind power increases with $L_{\rm{AGN}}$, and so $v_{\rm{s}}$, and hence the post-shock temperature of the shocked ISM layer, also increases (see Fig.~\ref{RvSimFig}). It thus takes longer for $T_{\rm{s}}$ to decrease enough for radiative cooling to become efficient. 

In the bottom right panel of Fig.~\ref{TFig}, we see that the metallicity has no effect on $T_{\rm{s}}$ at early times, when radiative cooling of the swept up gas is inefficient. However, since the metal line cooling rate scales linearly with metallicity (see equation~\ref{line_cooling_eqn}), the higher metallicity runs are able to radiatively cool efficiently at a (slightly) higher temperature. Thus the cooling time decreases with increasing metallicity. 

\begin{figure}
\centering
\mbox{
	\includegraphics[width=84mm]{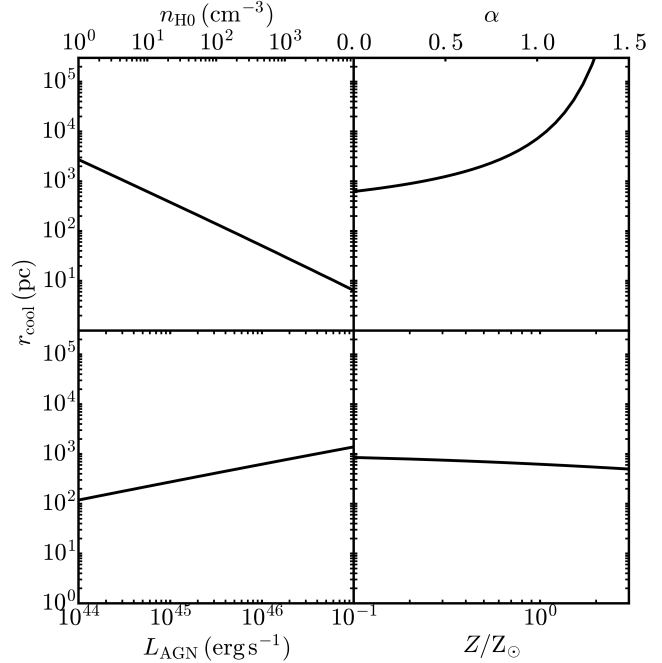}}
\caption{Cooling radius, $r_{\rm{cool}}$, of the outer shell of swept up gas, plotted against $n_{\rm{H}0}$ (top left), $\alpha$ (top right), $L_{\rm{AGN}}$ (bottom left), and $Z$ (bottom right). The cooling radius decreases with increasing $n_{\rm{H}0}$ and $Z$, and with decreasing $\alpha$ and $L_{\rm{AGN}}$.} 
\label{rCoolFig}
\end{figure}

We can also use the analytic model to calculate the cooling radius, $r_{\rm{cool}}$, at which the swept up gas in the outer shell of the outflow first cools to $10^{4} \, \rm{K}$. Fig.~\ref{rCoolFig} shows $r_{\rm{cool}}$ plotted against $n_{\rm{H}0}$, $\alpha$, $L_{\rm{AGN}}$ and $Z$ in the top left, top right, bottom left and bottom right panels, respectively. In each panel, all parameters are fixed at their fiducial values except the parameter that is being varied. 

As $n_{\rm{H}0}$ increases from 1 to $10^{4} \, \rm{cm}^{-3}$, $r_{\rm{cool}}$ decreases from $2700$ to $6 \, \rm{pc}$ (top left panel). At higher densities, the outflow is slower (Fig.~\ref{RvSimFig}) and the shocked ISM layer cools quicker (Fig.~\ref{TFig}), hence it can cool at smaller radii. In the top right panel, we see that $r_{\rm{cool}}$ increases with increasing $\alpha$. This is because, as noted above, a steeper density profile results in lower densities at radii $>$$100 \, \rm{pc}$, which reduces the radiative cooling rate. 

In the bottom left panel, $r_{\rm{cool}}$ increases from $100$ to $1400 \, \rm{pc}$ as $L_{\rm{AGN}}$ increases from $10^{44}$ to $10^{47} \, \rm{erg} \, \rm{s}^{-1}$. This is because both $v_{\rm{s}}$ and the cooling time increase with increasing $L_{\rm{AGN}}$ (Figs.~\ref{RvSimFig} and \ref{TFig}, respectively). 

We see in the bottom right panel that $r_{\rm{cool}}$ decreases slowly with increasing metallicity. This is due to the decrease in the cooling time as metallicity increases (Fig.~\ref{TFig}), while $v_{\rm{s}}$ is unchanged (Fig.~\ref{RvSimFig}). 

\begin{figure}
\centering
\mbox{
	\includegraphics[width=84mm]{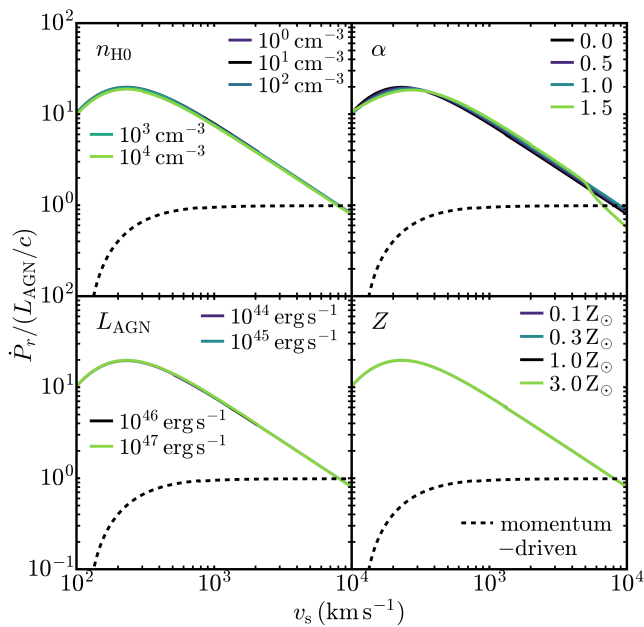}}
\caption{Momentum boost factor, $\dot{P}_{r} / (L_{\rm{AGN}} / c)$, versus forward shock velocity, $v_{\rm{s}}$, calculated from the analytic model with different $n_{\rm{H}0}$ (top left), $\alpha$ (top right), $L_{\rm{AGN}}$ (bottom left), and $Z$ (bottom right). As we saw in Fig.~\ref{PrDotSimFig}, the momentum boost factor at high $v_{\rm{s}}$ (i.e. early times) scales with $v_{\rm{s}}^{-1}$, as expected for an energy-driven wind, but it turns over at low $v_{\rm{s}}$. This relation is insensitive to the four parameters varied here, except for the high-$\alpha$ model in the top right panel. The black dashed curves show a momentum-driven model with our fiducial parameters, which is lower than the energy-driven models, by more than an order of magnitude at late times.}
\label{PrDotFig}
\end{figure} 

Fig.~\ref{PrDotFig} shows how the momentum boost factor (i.e. the instantaneous rate of change of radial momentum of the outflow normalised by the momentum injection rate of the AGN, $\dot{P}_{r} / (L_{\rm{AGN}} / c)$) varies with $v_{\rm{s}}$ in the analytic models for different $n_{\rm{H}0}$ (top left), $\alpha$ (top right), $L_{\rm{AGN}}$ (bottom left), and $Z$ (bottom right). As we saw in Fig.~\ref{PrDotSimFig}, the momentum boost factor scales with $v_{\rm{s}}^{-1}$ at high velocities, as we would expect for an energy-conserving wind, but it peaks at $v_{\rm{s}} \approx 200 \, \rm{km} \, \rm{s}^{-1}$ with a maximum momentum boost of $\approx 20$. The relation between the momentum boost factor and $v_{\rm{s}}$ is independent of $n_{\rm{H}0}$, $L_{\rm{AGN}}$ and $Z$. The only deviations in this relation are seen in the high-$\alpha$ model ($\alpha = 1.5$), where the momentum boost factor is slightly lower than the other models at high velocities ($v_{\rm{s}} \ga 5000 \, \rm{km} \, \rm{s}^{-1}$). However, we caution that these deviations in the high-$\alpha$ model are likely to be unphysical, and may be a consequence of the fact that the assumptions in our analytic break down at small radii. In particular, these deviations are sensitive to the initial radius that we start integrating from (which we take to be $0.1 \, \rm{pc}$ by default). 

\begin{figure}
\centering
\mbox{
	\includegraphics[width=84mm]{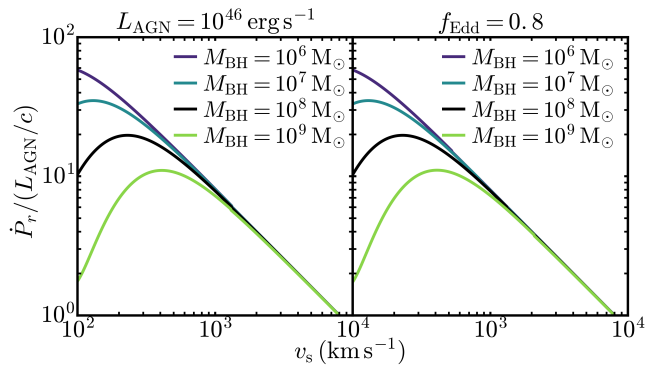}}
\caption{Momentum boost factor, $\dot{P}_{r} / (L_{\rm{AGN}} / c)$, versus forward shock velocity, $v_{\rm{s}}$, from the analytic model for variations in the black hole mass and host galaxy potential (with velocity dispersion $\sigma$ following the $M_{\rm{BH}} - \sigma$ relation from \citealt{gultekin09}). The left and right hand panels use a fixed AGN luminosity $L_{\rm{AGN}} = 10^{46} \, \rm{erg} \, \rm{s}^{-1}$ and a fixed Eddington ratio $f_{\rm{Edd}} = 0.8$, respectively. As the velocity dispersion of the gravitational potential increases, the peak in the momentum boost factor decreases and moves to higher $v_{\rm{s}}$, as more energy is lost from the outflow due to work done against the gravitational potential.} 
\label{PrDotBHFig}
\end{figure}

The turn-over in this relation at low $v_{\rm{s}}$ is due to the gravitational potential, which is dominated by the host galaxy at radii $\ga 5 \, \rm{pc}$ for our fiducial parameters. This turn-over occurs at $v_{\rm{s}} \sim \sigma$, where $\sigma$ is the velocity dispersion of the host galaxy potential. To demonstrate that this is the case, we show in Fig.~\ref{PrDotBHFig} the relation between the momentum boost factor and $v_{\rm{s}}$ from the analytic model for different gravitational potentials of the black hole and host galaxy. We consider a range of black hole masses from $10^{6}$ to $10^{9} \, \rm{M}_{\odot}$, and we take the velocity dispersion of the host galaxy potential, $\sigma$ (as used in equation~\ref{mgal_potential_eqn}), from the $M_{\rm{BH}} - \sigma$ relation of \citet{gultekin09}. In the left panel of Fig.~\ref{PrDotBHFig}, we use a fixed AGN luminosity of $10^{46} \, \rm{erg} \, \rm{s}^{-1}$ as we vary the black hole mass, while in the right panel we use a fixed Eddington ratio of 0.8, as used in the fiducial model. 

As we increase the velocity dispersion of the gravitational potential, the maximum momentum boost factor decreases, and it peaks at a higher $v_{\rm{s}}$. Thus the deviations from the simple energy-conserving relation ($\propto v_{\rm{s}}^{-1}$) arise because energy is lost from the outflow due to work done against the gravitational potential. The left and right hand panels of Fig.~\ref{PrDotBHFig} are identical, which is consistent with the lower left panel of Fig.~\ref{PrDotFig}, where we saw that this relation is insensitive to the AGN luminosity. 

The analytic model is always in the `energy-driven' regime for the range of parameters that we consider here, in the sense that the hot shocked wind bubble remains hot and its thermal pressure drives the outflow. The opposite limit would be a `momentum-driven' outflow, in which the shocked wind bubble cools rapidly. Note that, while this latter scenario is sometimes called a `momentum-conserving' outflow, the momentum boost factor is not necessarily equal to unity at all times, as momentum can still be lost due to the gravitational potential. To illustrate the differences between energy- and momentum-driven outflows, we also show in Fig.~\ref{PrDotFig} the momentum boost factor from a modified analytic model evaluated in the momentum-driven limit (black dashed curve), using our fiducial parameters. We calculate this by replacing the thermal pressure term in equation~\ref{eqn_of_motion} with the momentum injection rate that is directly injected by the AGN, from equation~\ref{momentum_inj_eqn} (see also \citealt{king05}). This model initially has a momentum boost of unity, at high $v_{\rm{s}}$, which decreases at $v_{\rm{s}} \la 500 \, \rm{km} \, \rm{s}^{-1}$ due to gravity. 

In Figs.~\ref{PrDotFig} and \ref{PrDotBHFig}, we calculated the momentum boost factor directly from the analytic model using the instantaneous rate of change of the radial momentum of the outflow. However, we will show in Section~\ref{obs_boost_sect} that this is not necessarily equivalent to observational estimates, which infer a time-averaged momentum boost factor from the size, mass and velocity of the outflow. 

\section{Comparison with observations}\label{obs_sect} 

\subsection{Molecular outflow rates}

From Figs.~\ref{TFig} and \ref{rCoolFig}, we see that, for a wide range of parameters, the shocked ISM layer is able to cool within $1 \, \rm{Myr}$ and at radii below $\sim 1 \, \rm{kpc}$, as required by the spatial extent and flow times of molecular outflows observed in luminous quasars \citep[e.g.][]{gonzalezalfonso17}. In our simulations from Paper~\textsc{i}, we found that, once the shocked ISM layer has cooled, it can rapidly form molecules, assuming that dust grains are present in the outflow. We can then use these results from Paper~\textsc{i} together with the analytic model to make predictions for how the molecular mass outflow rates and velocities vary with the physical parameters, which we can compare to observations. This will enable us to test the predictions for our molecular outflow simulations, and hence the assumptions that go into these simulations such as the presence of dust grains to catalyse molecule formation in the outflow, across a much wider range of physical parameters than we could with the simulations alone. We note that other models involving the entrainment of existing molecular clouds could have very different efficiencies for accelerating molecular gas to the velocities of observed AGN-driven molecular outflows, e.g. if the cross section of pre-existing molecular clouds is small or due to destruction by hydrodynamical instabilities (see Section~\ref{intro_sect}). Therefore, the mass outflow rates and velocities predicted by our models are non-trivial tests of the in-situ molecule formation scenario. 

We showed in Paper~\textsc{i} that, in the fiducial simulation (nH10\_L46\_Z1), the molecular fraction of outflowing gas was $f_{\rm{H}_{2}} = M_{\rm{H}_{2}} / M_{\rm{H}_{tot}} = 0.2$ after $1 \, \rm{Myr}$. This increased slightly in the low-luminosity simulation, to $f_{\rm{H}_{2}} = 0.3$. For the analytic model, we therefore assume that, after the shocked ISM layer has cooled to $10^{4} \, \rm{K}$, 20 per cent of its hydrogen mass is molecular, i.e. that $M_{\rm{H}_{2}} = f_{\rm{H}_{2}} X_{\rm{H}} M_{\rm{s}}$, where $f_{\rm{H}_{2}} = 0.2$, and $X_{\rm{H}} = 0.7$ is the hydrogen mass fraction. In the simulations, the assumption $f_{\rm{H}_{2}} = 0.2$ only holds at solar metallicity, as it decreased by more than an order of magnitude in the low-metallicity run ($0.1 \, \rm{Z}_{\odot}$) at $1 \, \rm{Myr}$. Also, we do not know how the H$_{2}$ fraction after the shocked ISM layer has cooled will depend on $n_{\rm{H}0}$, because the low-density simulation did not cool within $1 \, \rm{Myr}$. However, at densities higher than the fiducial run we expect $f_{\rm{H}_{2}}$ to be higher, as high densities are more conducive to molecule formation. Assuming $f_{\rm{H}_{2}} = 0.2$ at solar metallicity therefore gives a lower limit on $M_{\rm{H}_{2}}$, although at high densities we can underestimate it by no more than a factor of 5, as $f_{\rm{H}_{2}}$ cannot exceed unity. 

\begin{figure}
\centering
\mbox{
	\includegraphics[width=84mm]{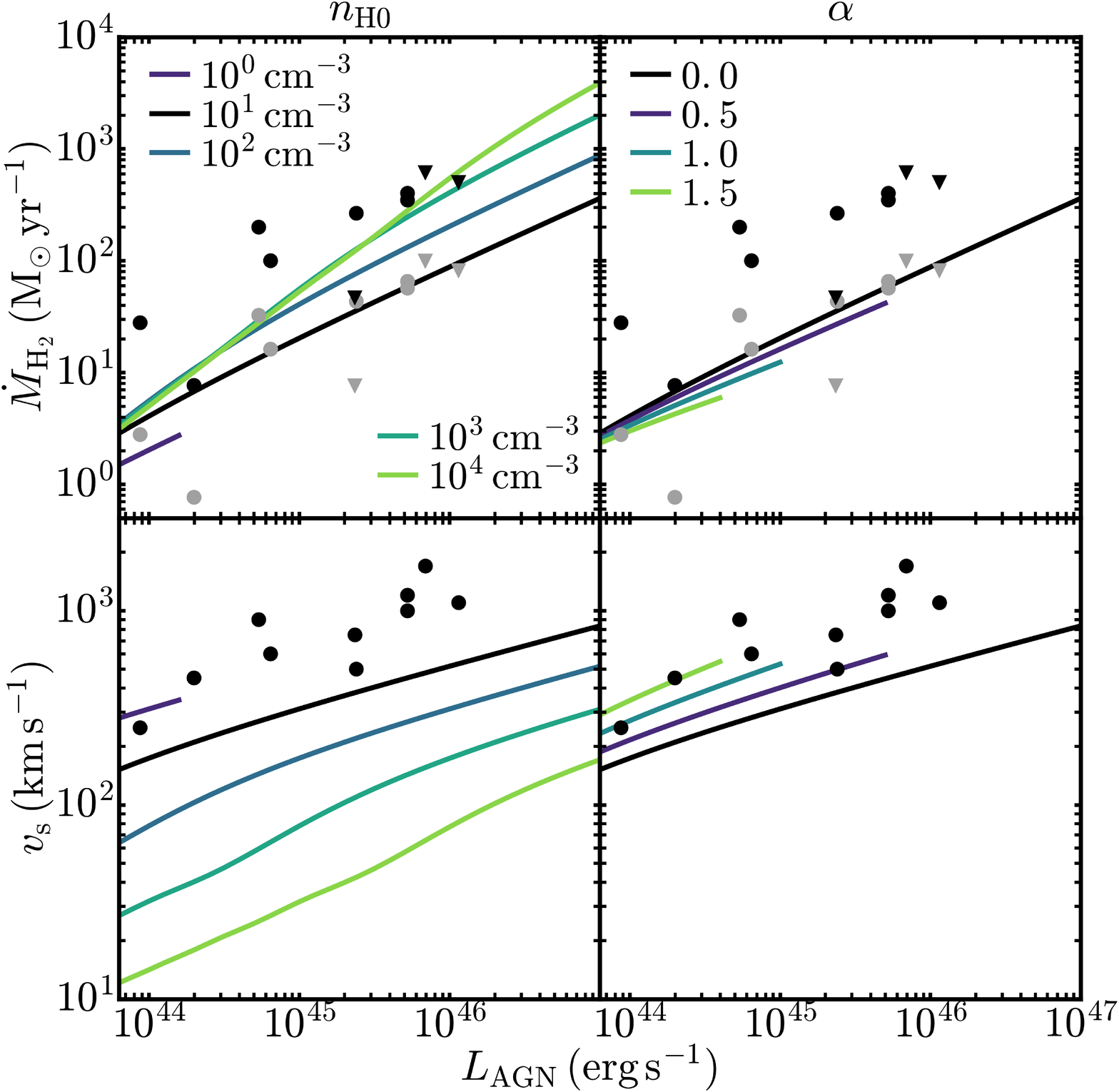}}
\caption{Mass outflow rate of H$_{2}$ (top row) and outflow velocity (bottom row) plotted against AGN luminosity. Solid curves are calculated at $1 \, \rm{Myr}$ in the analytic model with different values of $n_{\rm{H}0}$ (left-hand column) and $\alpha$ (right-hand column), while symbols show detections (circles) and upper limits (triangles) from observed AGN host galaxies in the extended sample of \citet{cicone14}. In the top row, black symbols show outflow rates calculated using the same CO to H$_{2}$ conversion factor assumed by \citet{cicone14}, while grey symbols show the observed outflow rates using the conversion factors calculated from our simulations in Paper~\textsc{i}, which are lower by a factor $\approx 5-10$, depending on which CO line was used for each individual galaxy. In the bottom row, the observed velocities are maximum line of sight velocities measured from CO spectra. The analytic models with $n_{\rm{H}0} = 10 - 100 \, \rm{cm}^{-3}$ reproduce the observed outflow rates using the lower conversion factors, while densities of at least $n_{\rm{H}0} = 10^{3} - 10^{4} \, \rm{cm}^{-3}$ are needed to reproduce the observations with the higher conversion factor. The analytic models underpredict the observed maximum velocities, which may be due to the assumption of a uniform medium in our models (see text).}
\label{MH2DotFig}
\end{figure} 

The top row of Fig.~\ref{MH2DotFig} shows the mass outflow rate of H$_{2}$, which is calculated from the analytic model as: 

\begin{equation}\label{MH2_dot_eqn} 
\dot{M}_{\rm{H}_{2}} = \frac{M_{\rm{H}_{2}}}{t_{\rm{flow}}} = \frac{f_{\rm{H}_{2}} X_{\rm{H}} M_{\rm{s}} v_{\rm{s}}}{R_{\rm{s}}}, 
\end{equation} 
where $f_{\rm{H_{2}}} = 0.2$ if $T_{\rm{s}} \leq 10^{4} \, \rm{K}$, or zero otherwise. The outflow rates are calculated after $1 \, \rm{Myr}$, and are plotted against $L_{\rm{AGN}}$ for different values of $n_{\rm{H}0}$ (left panel) and $\alpha$ (right panel). We consider only solar metallicity here, as the metallicity dependence of $f_{\rm{H}_{2}}$ is uncertain. This is also more relevant for the observed molecular outflows in luminous quasars, which are typically found in ULIRGs with metallicities close to solar \citep{rupke08, kilercieser14}. 

The analytic models are shown by the solid curves, while the symbols show observed ULIRGs from \citet{cicone14}. We show only galaxies in their sample identified as Seyfert 1 or 2 galaxies, since not all of the galaxies in their sample host luminous quasars. Their extended sample includes 6 AGN host galaxies observed by \citet{cicone14}, plus a further 4 taken from the literature \citep{wiklind95, maiolino97, cicone12, feruglio13a, feruglio13b}. \citet{cicone14} calculated the outflow rate as $\dot{M}_{\rm{H}_{2}} = 3 M_{\rm{H}_{2}} v / R$, assuming that the outflow is uniformly distributed\footnote{However, \citet{gonzalezalfonso17} note that, for a volume-filling wind to have a steady flow with constant velocity, we expect the density at the outer radius of the outflow to be 1/3 that of the average density, which would cancel the additional factor of 3 in \citet{cicone14}'s estimate for $\dot{M}_{\rm{H}_{2}}$.} over a radius $R$. However, as our model is an outflowing shell, this is a factor 3 larger than we use in equation~\ref{MH2_dot_eqn}. We therefore divide the outflow rates from \citet{cicone14} by a factor of 3. The circles in the top row of Fig.~\ref{MH2DotFig} show detections from \citet{cicone14}, while triangles show upper limits. The black symbols show the outflow rates reported by \citet{cicone14}, divided by a factor of 3. These were calculated assuming a CO to H$_{2}$ conversion factor of $\alpha_{\rm{CO}} = 0.8 \, \rm{M}_{\odot} \, (\rm{K} \, \rm{km} \, \rm{s}^{-1} \, \rm{pc}^{2})^{-1}$. Some of the outflows in their sample were measured from the CO 1$-$0 line, while others used the 2$-$1 line; they used the same conversion factor for both lines. However, in Paper~\textsc{i} we found that, in our fiducial simulation (nH10\_L46\_Z1), the CO to H$_{2}$ conversion factors for the 1$-$0 and 2$-$1 lines were $\alpha_{\rm{CO}} = 0.13$ and $0.08 \, \rm{M}_{\odot} \, (\rm{K} \, \rm{km} \, \rm{s}^{-1} \, \rm{pc}^{2})^{-1}$, respectively. The grey symbols in the top panel of Fig.~\ref{MH2DotFig} show the outflow rates that we would get from \citet{cicone14} if we instead used the conversion factors from our fiducial simulation, corresponding to the given line used for each individual observation. 

In both panels in the top row of Fig.~\ref{MH2DotFig}, $\dot{M}_{\rm{H}_{2}}$ increases with increasing $L_{\rm{AGN}}$, with a similar slope as in the observations. $\dot{M}_{\rm{H}_{2}}$ increases slowly with increasing $n_{\rm{H}0}$, by a factor $\approx 10$ as $n_{\rm{H}0}$ increases from 10 to $10^{4} \, \rm{cm}^{-3}$ at $L_{\rm{AGN}} = 10^{47} \, \rm{erg} \, \rm{s}^{-1}$. The analytic models at $n_{\rm{H}0} = 10 - 100 \, \rm{cm}^{-3}$ agree well with the observations from \citet{cicone14} using the $\alpha_{\rm{CO}}$ conversion factor from the simulations (grey symbols). However, only the highest density models, at $n_{\rm{H}0} \approx 10^{3} - 10^{4} \, \rm{cm}^{-3}$, are able to reach the observed outflow rates from \citet{cicone14} using their original conversion factor (black symbols). We again note that, in the high-density models, we may underestimate $f_{\rm{H}_{2}}$, and hence $\dot{M}_{\rm{H}_{2}}$, by up to a factor of 5. In the top right panel of Fig.~\ref{MH2DotFig}, we see that $\dot{M}_{\rm{H}_{2}}$ depends only weakly on $\alpha$, when the density profile is normalised at $100$ pc. However, the high-$\alpha$ models are only able to form molecules within 1 Myr at low AGN luminosities. 

The bottom row of Fig.~\ref{MH2DotFig} shows the outflow velocity after $1 \, \rm{Myr}$ in the analytic model (solid curves) and the maximum line of sight velocity measured from CO spectra in the Seyfert 1 and 2 AGN host galaxies in the extended sample from \citet{cicone14} (black circles), plotted against $L_{\rm{AGN}}$ for different $n_{\rm{H}0}$ (left panel) and $\alpha$ (right panel). The solid curves are only shown for models that have cooled to $T_{\rm{s}} \leq 10^{4} \, \rm{K}$ after $1 \, \rm{Myr}$. 

As $n_{\rm{H}0}$ increases, $v_{\rm{s}}$ decreases. Only the lowest density models, with $n_{\rm{H}0} = 1 \, \rm{cm}^{-3}$, reproduce the observed maximum velocities. However, at such a low density, molecular outflows can only form within 1 Myr at $L_{\rm{AGN}} \la 2 \times 10^{44} \, \rm{erg} \, \rm{s}^{-1}$, because at higher AGN luminosities they cannot cool within that time. In the bottom right panel, the outflow velocity increases with increasing $\alpha$. The $\alpha = 1.5$ model agrees well with the observed velocities. However, this model only forms H$_{2}$ within 1 Myr at AGN luminosities $L_{\rm{AGN}} \la 4 \times 10^{44} \, \rm{erg} \, \rm{s}^{-1}$, so it still cannot explain the high velocities observed at higher AGN luminosities. 

We also found that the simulations tend to underpredict the observed velocities (see fig.~8 of Paper~\textsc{i}). As we noted in Paper~\textsc{i}, it is possible that the low velocities that we find in the simulations and the analytic model may be because we do not include density inhomogeneities in the ambient ISM. In the presence of inhomogeneities, the maximum velocity will be determined by gas escaping through low-density channels, while the bulk of the outflowing H$_{2}$ mass may be along paths at higher densities. 

\subsection{Momentum boost factors}\label{obs_boost_sect}  

\begin{figure}
\centering
\mbox{
	\includegraphics[width=84mm]{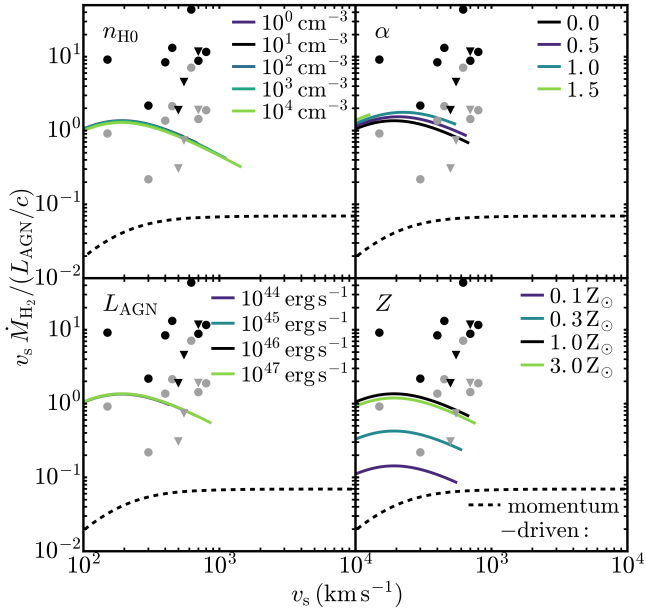}}
\caption{Observational estimates of the momentum boost factor versus $v_{\rm{s}}$ from the analytic model (solid curves) for different $n_{\rm{H}0}$ (top left), $\alpha$ (top right), $L_{\rm{AGN}}$ (bottom left), and $Z$ (bottom right), and from the observational sample of ULIRGs that host AGN from \citet{cicone14} (circles and triangles, indicating detections and upper limits respectively) using the same CO to H$_{2}$ conversion factor from that work (black symbols), and using the conversion factor from our simulations in Paper~\textsc{i} (grey symbols). The black dashed curves show a modified analytic model evaluated in the momentum-driven limit. The observational estimates, which are averaged over the flow time ($R_{\rm{s}} / v_{\rm{s}}$), for the energy-conserving analytic model are a factor $\approx 20$ lower than the instantaneous momentum boost factors shown in Fig.~\ref{PrDotFig}. This is partly because we include only the molecular component here, and partly due to discrepancies between the two different estimates for $\dot{P}_{r}$.}
\label{PrDotObsFig}
\end{figure} 

In Figs.~\ref{PrDotFig} and \ref{PrDotBHFig}, we calculated the instantaneous rate of change of radial momentum of the shocked ISM layer, $\dot{P}_{r}$, directly from the analytic model, by measuring the change in the radial momentum in each time-step and dividing by the length of the time-step. However, observations of molecular outflows typically infer a momentum outflow rate $\dot{P}_{r}^{\rm{obs}}$ from the mass, velocity and radius of the outflow, with $\dot{P}_{r}^{\rm{obs}} = v_{\rm{s}} \dot{M}_{\rm{H}_{2}}$. To compare the momentum boost factors from the analytic model to observations, we calculated an observational estimate for $\dot{P}_{r}^{\rm{obs}}$ in the same way, using the H$_{2}$ mass outflow rates calculated as in equation~\ref{MH2_dot_eqn}. The resulting momentum boost factors from the analytic model are shown in Fig.~\ref{PrDotObsFig}, plotted against $v_{\rm{s}}$ (solid curves). For the runs at different metallicities, we assumed that the H$_{2}$ fraction, $f_{\rm{H}_{2}}$, scales linearly with metallicity, which is approximately what we found for the simulations with varying metallicity in Paper \textsc{i}. Above solar metallicity, we capped $f_{\rm{H}_{2}}$ at 0.2, as a conservative estimate. However, this scaling with metallicity is highly uncertain. Note that the trends with metallicity seen in the lower right panel of Fig.~\ref{PrDotObsFig} are primarily driven by this uncertain scaling of $f_{\rm{H}_{2}}$ with metallicity. 

The black symbols in Fig.~\ref{PrDotObsFig} show the observational sample of ULIRGs that host an AGN from \citet{cicone14} using the same CO to H$_{2}$ conversion factor used in that work and dividing by a factor of 3 to account for the different assumed geometries of the outflow (circles indicate detections, triangles indicate upper limits). The grey symbols show the observational sample of \citet{cicone14}, but using the CO to H$_{2}$ conversion factor calculated from our fiducial simulation (nH10\_L46\_Z1) in Paper \textsc{i}, which reduces $M_{\rm{H}_{2}}$, and hence the momentum boost factor, by a factor of $\approx 5-10$, depending on which CO line was used. 

Compared to Fig.~\ref{PrDotFig}, we see that the observational estimates of the momentum boost factor from the analytic model are lower by a factor $\approx 20$. This is partly because the observational estimate (as defined here) uses only the molecular component of the outflow, with a mass that is a factor $f_{\rm{H_{2}}} X_{\rm{H}} = 0.14$ times the total mass of the outflow. However, as we noted above, our assumption that $f_{\rm{H}_{2}} = 0.2$ at solar metallicity is likely to underestimate the H$_{2}$ mass, and hence the momentum boost factor of the molecular component, by up to a factor of 5 in the high density analytic models. 

The molecular fraction still does not fully explain the difference between the instantaneous (Fig.~\ref{PrDotFig}) and observational (Fig.~\ref{PrDotObsFig}) estimates of the momentum boost factors in the analytic model. If we differentiate the radial momentum of the outflow, we see that the instantaneous $\dot{P}_{r}$ can be expressed as: 

\begin{equation}\label{pr_phys_eqn}
\dot{P}_{r} = \dot{M}_{\rm{s}} v_{\rm{s}} + M_{\rm{s}} \dot{v}_{\rm{s}}. 
\end{equation} 
If we take $f_{\rm{H}_{2}} X_{\rm{H}} = 1$, i.e. the entire mass of the outflow is in molecular hydrogen, then the observational estimate $\dot{P}_{r}^{\rm{obs}}$ is: 

\begin{align}\label{pr_obs_eqn} 
\dot{P}_{r}^{\rm{obs}} &= \dot{M}_{\rm{s}}^{\rm{obs}} v_{\rm{s}} \nonumber \\ 
 &= \frac{M_{\rm{s}}^{\rm{obs}} v_{\rm{s}}}{R_{\rm{s}} / v_{\rm{s}}} \nonumber \\ 
 &= \frac{M_{\rm{s}} v_{\rm{s}}^{2}}{R_{\rm{s}}}. 
\end{align}
From the second line in equation~\ref{pr_obs_eqn}, we see that the observational estimate is equivalent to the observed momentum in the outflow divided by the flow time, $R_{\rm{s}} / v_{\rm{s}}$. In other words, the observational estimate is the rate of change of the outflow momentum averaged over the flow time. 

There are two discrepancies between these two estimates of $\dot{P}_{r}$. Firstly, the observational estimate does not include the second term in the right hand side of equation~\ref{pr_phys_eqn}, which arises from the acceleration of the outflow. We note that FGQ12 also neglected this second term when calculating the momentum boost factors, for example in their Fig. 4. However, since the outflow is decelerating, this term will be negative, and so the observational estimate would tend to be higher than the instantaneous value that we calculate from the models, which is the opposite of what is seen in Figs.~\ref{PrDotFig} and \ref{PrDotObsFig}. In the analytic model, we find that the magnitude of this term is typically $\approx 30$ per cent of $\dot{P}_{r}$ at $v_{\rm{s}}$ above the peak in the momentum boost factor, although it becomes comparable to $\dot{P}_{\rm{r}}$ at $v_{\rm{s}}$ below the peak. 

Secondly, the instantaneous rate of change of mass, $\dot{M}_{\rm{s}}$, in equation~\ref{pr_phys_eqn} is the rate at which mass is being added to the shocked ISM reservoir, as the outflow sweeps up mass from the ambient ISM. However, the observational estimate of the rate of change of mass, $\dot{M}_{\rm{s}}^{\rm{obs}}$, in equation~\ref{pr_obs_eqn} is the rate at which the shocked ISM mass is outflowing. These two definitions are not equivalent. This discrepancy accounts for the remaining differences between the two estimates of the momentum boost factor in the analytic model. 

The momentum boost factor is often used as a diagnostic to distinguish between energy- and momentum-driven outflows. For example, \citet{stern16} compiled several observational estimates of momentum boost factors $\gg 1$ for galaxy-scale outflows in luminous quasars, which suggest energy-conserving outflows. However, some recent observational studies of molecular outflows have reported lower momentum boosts. \citet{feruglio17} and \citet{veilleux17} measured momentum boost factors of $\approx 2-6$ and $\approx 1.5-3$, respectively, in molecular outflows from luminous AGN. They defined the momentum boost as in our equation~\ref{pr_obs_eqn}, although \citet{feruglio17} included an additional factor of 3 due to the assumed geometry. However, we have seen that observed values close to unity may still be consistent with our energy-driven models, at least for our assumed molecular fraction and host galaxy potential. These results highlight that it is critical to use consistent definitions when comparing observed momentum boost factors with theoretical predictions. 

To quantify how these observational estimates can distinguish between energy- and momentum-driven outflows, we also show a modified analytic model evaluated in the momentum-driven limit (as described in Section.~\ref{param_sect}) in Fig.~\ref{PrDotObsFig}. Note that, in this model, we do not self-consistently determine whether or not the swept up gas has cooled, so we plot it for all $v_{\rm{s}}$, using the same molecular fraction as for the energy-driven models. We see that the momentum-driven model is approximately an order of magnitude below the energy-driven models. Therefore, if we can measure all phases (molecular, neutral atomic and ionic) in an observed AGN outflow, along with the host galaxy potential, then the momentum boost factor can still be used to distinguish between energy- and momentum-driven outflows. For comparison, \citet{rupke17} compile outflow measurements of ionized, neutral atomic and molecular gas in nearby quasars. They find total momentum boost factors (summed over all observed gas phases in the outflow) of 0.01$-$20. The dashed curve in Fig.~\ref{PrDotObsFig} shows only the momentum of the molecular component in our momentum-driven model. If we divide this curve by the assumed molecular fraction, we see that, summed over all gas phases, momentum boosts below 0.5 (at $v_{\rm{s}} \ga 400 \, \rm{km} \, \rm{s}^{-1}$) are consistent with a momentum-driven scenario. Half of the type 1 quasars in \citet{rupke17} have momentum boost factors below 0.5, while the other half have higher momentum boosts favoring energy-conserving models. We stress, however, that observationally-inferred mass outflow rates (and therefore momentum boosts) can be highly uncertain, so observations cannot yet robustly distinguish energy vs. momentum-conserving scenarios.

The observational estimates of the momentum boost factor from the energy-driven analytic models at solar metallicity in Fig.~\ref{PrDotObsFig} are comparable to the observations of \citet{cicone14} if we use the CO to H$_{2}$ conversion factor from the simulations (grey symbols), although the models do not reproduce the observed scatter. However, as noted above, the high-density models are likely to underpredict the H$_{2}$ mass by up to a factor of 5. Since the observed systems will likely span a range of densities, this may explain why the models do not reproduce the observed scatter. 

\section{Dust formation and destruction in AGN-driven winds}\label{dust_sect} 

In the previous section, we used the results of our simulations from Paper~\textsc{i} to argue that, once the shocked ISM layer has cooled to $10^{4} \, \rm{K}$, we expect 20 per cent of the hydrogen mass to be molecular. However, a major caveat of these simulations is that we assumed a Milky Way dust-to-metals ratio. A high dust abundance is important for molecule formation, as dust grains catalyse the formation of H$_{2}$ and help to shield molecules from dissociating radiation. Indeed, we showed in Paper~\textsc{i} that, if we reduce the dust abundance by a factor of 100, the resulting H$_{2}$ outflow rate after $1 \, \rm{Myr}$ is reduced by a factor of 150 (see the lowDust100 model in fig.~6 of Paper~\textsc{i}). However, it is currently unclear whether dust grains will be able to survive in an AGN wind. For example, using simulations of the shocked ISM layer of an AGN wind, \citet{ferrara16} found that dust grains can be rapidly destroyed by sputtering due to the high gas temperatures, although their models did not include dust formation mechanisms such as the accretion of metals from the gas phase on to grains after the shocked ISM layer has cooled. Also, new dust grains from the ambient medium or star formation-driven outflows may be mixed into the AGN wind. Dust has been observed in star formation-driven galactic winds \citep[e.g.][]{hoopes05, roussel10, melendez15}. Additionally, \citet{barcosmunoz18} recently detected dust continuum emission that is spatially coincident with a molecular outflow in Arp 220, which may be driven by star formation or an AGN.

To explore whether our assumption of a Milky Way dust-to-gas ratio is feasible, we used our analytic AGN wind model to track the formation and destruction of dust grains in the shocked ISM layer. Our dust model is loosely based on the models of \citet{asano13}, who investigated dust evolution in galaxies. In particular, we have adapted their equation~4 for the evolution of the dust mass as follows: 

\begin{equation}\label{dust_eqn}  
\frac{\rm{d}\mathit{M}_{\rm{d}}}{\rm{d}\mathit{t}} = \epsilon(T_{\rm{s}}) \mathcal{D}_{\rm{MW}} \frac{\rm{d}\mathit{M}_{\rm{s}}}{\rm{d}\mathit{t}} - \frac{M_{\rm{d}}}{\tau_{\rm{sput}}} + \frac{M_{\rm{d}}}{\tau_{\rm{acc}}} \left( 1 - \frac{M_{\rm{d}}}{Z M_{\rm{s}}} \right), 
\end{equation} 
where $M_{\rm{d}}$ is the dust mass in the shocked ISM layer, $\epsilon (T_{\rm{s}})$ parameterises the fraction of dust swept up from the ambient medium that is mixed into the shocked ISM layer (as discussed further below), and $\mathcal{D}_{\rm{MW}} = 6.3 \times 10^{-3}$ is the Milky Way dust-to-gas ratio at solar metallicity $\rm{Z}_{\odot} = 0.0129$. The time-scales for dust destruction via sputtering ($\tau_{\rm{sput}}$) and dust growth via the accretion of metals ($\tau_{\rm{acc}}$) are defined below. 

Compared to the model of \citet{asano13}, we do not include the loss of dust grains locked up in newly-formed stars (as we do not follow star formation in the wind; we do not expect that this would have a significant impact on the dust content in the wind), the yield of dust grains from stars has been replaced by the injection of grains mixed into the shocked ISM layer from newly swept-up gas, and destruction by supernovae has been replaced with destruction by sputtering from the AGN wind. Furthermore, \citet{asano13} multiply the accretion term by a parameter $\eta$, which is the mass fraction of the gas that is in cold clouds, where accretion can proceed. However, since we are interested in the dust-to-gas ratio in the cold phase, where molecules can form, we set $\eta$ to unity. 

The sputtering time-scale, $\tau_{\rm{sput}}$, can be calculated from equations~14 and 15 of \citet{tsai95} (see also equation~14 of \citealt{hirashita15}): 

\begin{align}\label{sput_time_eqn}  
\tau_{\rm{sput}} = 7.1 \times 10^{3} & \left(\frac{a}{0.1 \, \rm{\mu m}} \right) \left( \frac{n_{\rm{H}}}{10 \, \rm{cm}^{-3}} \right)^{-1} \nonumber \\ 
 & \times \left[ \left(\frac{2 \times 10^{6} \, \rm{K}}{T_{\rm{s}}} \right)^{2.5} + 1 \right] \, \rm{yr}, 
\end{align} 
where $a$ is the grain radius. Following \citet{ferrara16}, we assume $a = 0.1 \, \rm{\mu m}$, which is the average grain size for a \citet{mathis77} grain size distribution. 

We take the accretion time-scale, $\tau_{\rm{acc}}$, from equation~20 of \citet{asano13}: 

\begin{align}\label{acc_time_eqn} 
\tau_{\rm{acc}} = 2.2 \times 10^{4} & \left( \frac{a}{0.1 \, \mu\rm{m}} \right) \left( \frac{n_{\rm{H}}}{10^{4} \, \rm{cm}^{-3}} \right)^{-1} \nonumber \\  
 & \times \left( \frac{T}{10^{4} \, \rm{K}} \right)^{-1/2} \left( \frac{Z}{0.0129} \right)^{-1} \, \rm{yr}. 
\end{align} 
We again assume $a = 0.1 \, \rm{\mu m}$. We caution that this accretion time-scale is highly uncertain. In particular, it assumes a sticking coefficient of metals on to dust grains of unity, but this could be much lower at temperatures $\sim$10$^{4}$ K corresponding to the shocked ISM layer after it has cooled in our analytic model \citep[e.g.][]{zhukovska16}. Additionally, this neglects the effects of the strong UV radiation field from the AGN, which can positively charge the grains and thus further reduce the sticking coefficient \citep[e.g.][]{ferraraetal16}.

As the wind propagates outwards, it sweeps up gas at a rate of $\frac{\rm{d}\mathit{M}_{\rm{s}}}{\rm{d}\mathit{t}}$. This can be multiplied by $\mathcal{D}_{\rm{MW}}$ to obtain the rate at which dust is swept up from the ambient medium. However, not all of this dust will be mixed into the shocked ISM layer. Firstly, some of the dust grains will be destroyed as they first pass through the forward shock of the AGN wind. \citet{dwek96} calculated the mass fraction of grains that is destroyed in a fast non-radiative shock as a function of ambient ISM density and shock velocity (see their table~2). Based on these results, we assume that the mass fraction of dust that survives the initial shock is 0.45. This is the same value used by \citet{ferrara16} (also based on the results of \citealt{dwek96}), and is appropriate for a forward shock velocity of $1240 \, \rm{km} \, \rm{s}^{-1}$ and an ambient ISM density of $n_{\rm{H}} = 15 \, \rm{cm}^{-3}$, assuming an equal mixture of graphite and silicate grains. 

Secondly, after the dust grains have passed through the forward shock, they will continue to be destroyed due to sputtering by the hot shocked gas. At early times, when $T_{\rm{s}}$ is still high, we explicitly model this sputtering process (the second term in the right hand side of equation~\ref{dust_eqn}). However, once the shocked ISM layer has cooled, our analytic model treats the entire layer to be at the same temperature. This implies that dust grains swept up from the ambient ISM are immediately mixed into the cool phase of the shocked ISM layer. We can see from equation~\ref{sput_time_eqn} that, once $T_{\rm{s}} \la 10^{6} \, \rm{K}$, $\tau_{\rm{sput}}$ becomes very large and sputtering becomes inefficient, so this dust would not undergo further sputtering after passing through the forward shock in our analytic model. However, we find from our simulations that, when a parcel of gas is swept up from the ambient ISM after the bulk of the shocked ISM layer has cooled, this newly swept up material remains at the post-shock temperature for a finite period of time, during which the swept up dust will continue to be destroyed by sputtering, before it joins the cool phase of the shocked ISM layer. This additional period of sputtering is not explicitly included in our analytic model, which does not capture the multiphase nature of the shocked ISM layer. We therefore parameterise $\epsilon(T_{\rm{s}})$ in equation~\ref{dust_eqn} as follows: 

\begin{equation}\label{epsilon_dust_eqn} 
  \epsilon(T_{\rm{s}}) = 
  \begin{cases} 
    0.45 &T_{\rm{s}} \geq 10^{6} \, \rm{K} \\ 
    \epsilon_{\rm{mix}} &T_{\rm{s}} < 10^{6} \, \rm{K}, \\
  \end{cases} 
\end{equation} 
where $\epsilon_{\rm{mix}}$ is a free parameter that describes the fraction of newly swept up dust grains that are mixed into the cool phase of the shocked ISM layer after this layer has cooled below $10^{6} \, \rm{K}$, which is the temperature at which sputtering becomes inefficient (see equation~\ref{sput_time_eqn}). 

To consider what range of values of $\epsilon_{\rm{mix}}$ are feasible, we can compare the cooling time of freshly swept up gas to the sputtering time-scale. The post-shock temperature for an outflow velocity $v_{\rm{s}}$ is $T_{\rm{s}} \approx 3.33 \times 10^{6} \, \rm{K} \left( \frac{v_{\rm{s}}}{500 \, \rm{km} \, \rm{s}^{-1}} \right)^{2}$. At velocities $\la$1600 km s$^{-1}$ and solar metallicity, the cooling in the post-shock layer is dominated by metal line cooling, so we can estimate the cooling time, $t_{\rm{cool}}$, of the freshly swept up gas using equation~\ref{lambda_line_eqn}: 

\begin{equation} 
t_{\rm{cool}} = 1.3 \times 10^{4} \, \rm{yr} \left( \frac{\mathit{v}_{\rm{s}}}{500 \, \rm{km} \, \rm{s}^{-1}} \right)^{3.4} \left( \frac{\mathit{n}_{\rm{ambient}}}{10 \, \rm{cm}^{-3}} \right)^{-1} \left( \frac{\mathit{Z}}{\rm{Z}_{\odot}} \right)^{-1}, 
\end{equation} 
for $T_{\rm{s}} > 10^{5} \, \rm{K}$ ($v_{\rm{s}} \ga 100 \rm{km} \, \rm{s}^{-1}$), assuming that the post-shock density is four times the ambient density $n_{\rm{ambient}}$. Comparing this to the sputtering time-scale in equation~\ref{sput_time_eqn}, we find: 

\begin{align} 
\frac{\tau_{\rm{sput}}}{t_{\rm{cool}}} = 0.14 &\left( \frac{a}{0.1 \, \rm{\mu m}} \right) \left( \frac{Z}{Z_{\odot}} \right) \left( \frac{v_{\rm{s}}}{500 \, \rm{km} \, \rm{s}^{-1}} \right)^{-3.4} \\ \nonumber 
 & \times \left[ 0.28 \left( \frac{v_{\rm{s}}}{500 \, \rm{km} \, \rm{s}^{-1}} \right)^{-5} + 1\right]. 
\end{align} 

We can simplify this expression by noting that the `+1' term on the right-hand side only adds to this ratio, so that a lower bound is obtained by neglecting it: 

\begin{equation} 
\frac{\tau_{\rm{sput}}}{t_{\rm{cool}}} \ga 0.04 \left( \frac{a}{0.1 \, \rm{\mu m}} \right) \left( \frac{Z}{Z_{\odot}} \right) \left( \frac{v_{\rm{s}}}{500 \, \rm{km} \, \rm{s}^{-1}} \right)^{-8.4}. 
\end{equation} 

When this ratio exceeds unity, freshly swept up gas will cool before a significant fraction of the swept up dust grains can be sputtered. We see that, for solar metallicity and $v_{\rm{s}} = 500 \, \rm{km} \, \rm{s}^{-1}$, this is the case for large grains ($a \ga 2.5 \, \rm{\mu m}$). Furthermore, due to the strong scaling of this ratio with $v_{\rm{s}}$, average-sized grains ($a = 0.1 \, \rm{\mu m}$) are expected to survive sputtering in the shocked ISM layer at outflow velocities $v_{\rm{s}} \la 340 \, \rm{km} \, \rm{s}^{-1}$. Comparing to the lower panels of Fig.~\ref{MH2DotFig}, we see that, after 1 Myr, many of our analytic model runs are at these velocities. The observations from \citet{cicone14} find higher velocities than this (black points in Fig.~\ref{MH2DotFig}), although these are the maximum velocities that they measure from their CO spectra. In practice, we expect a range of outflow velocities as the outflow propagates through an inhomogeneous ambient medium. We therefore expect dust grains to survive along the low-velocity channels of the outflow. 

In addition to grains surviving a period of sputtering in the post-shock gas, there are also other mechanisms which may pollute the cold phase of the shocked ISM layers with dust grains to seed further grain growth. Firstly, as the outflow propagates through the host galaxy, dust grains in stellar winds from AGB stars and in star formation-driven outflows may mix into the AGN-driven outflow. As this material is swept up by the AGN outflow, it will be accelerated and shocked, which could sputter these dust grains as before. However, if some of the stellar winds and star formation-driven outflows are propagating in the same direction as the AGN outflow, the relative velocity between the two will be lower. We saw above that the ratio $\tau_{\rm{sput}} / t_{\rm{cool}}$ depends strongly on the shock velocity, so this could help alleviate the destruction of these grains. 

Secondly, when the outflow encounters a dense clump in the ambient medium, this clump may initially survive the forward shock as it passes over it. Then, as the fast outflow accelerates the clump, it will be shredded and destroyed by hydrodynamic instabilities \citep[e.g.][]{bruggen16}. Thus dense clumps in the ambient medium may be able to penetrate the hot layer of the swept up shell of the AGN outflow and inject its dust grains directly into the cold phase of this shell as it is shredded by the fast outflow. 

Once dust grains have entered the cold phase of the shocked ISM layer, either by surviving along the low-velocity channels, being injected by local stellar winds or galaxy-scale star formation-driven outflows, or being injected by dense clumps from the ambient medium penetrating the hot post-shock layer, they are likely to then mix throughout the cold phase by turbulence. Additionally, if the coupling between gas and dust grains is not perfect, this may futher enable mixing of grains throughout the cold phase. There are thus a number of mechanisms which may enable swept up grains to mix into the cold phase to seed further grain growth via accretion. However, the details of these mechanisms remain highly uncertain. We will therefore consider a wide range of values for the parameter $\epsilon_{\rm{mix}}$, from $10^{-6}$ to 0.45, to quantify how sensitive our results are to these uncertainties. 

\begin{figure}
\centering
\mbox{
	\includegraphics[width=84mm]{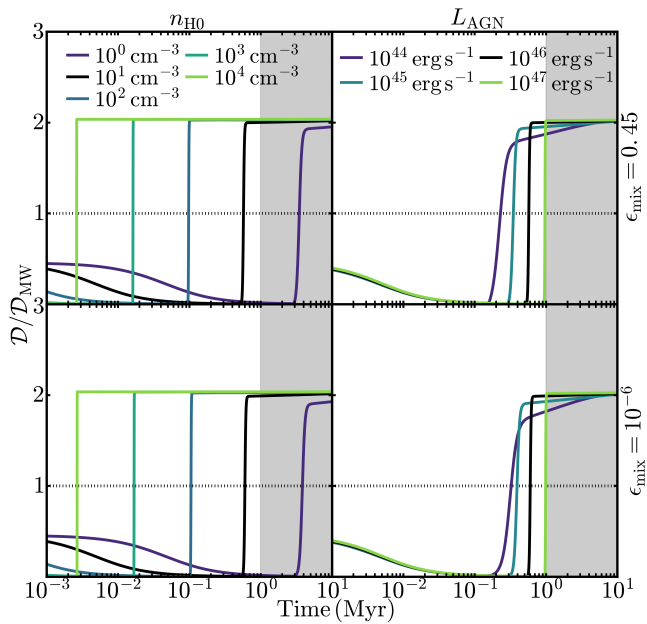}}
\caption{Evolution of the dust-to-gas mass ratio, $\mathcal{D} = M_{\rm{d}} / M_{\rm{s}}$, scaled to the Milky Way value, $\mathcal{D}_{\rm{MW}}$, from our analytic AGN wind model for different ambient ISM densities (left column) and AGN luminosities (right column) at solar metallicity. The parameters that are not varied in a given panel are held fixed at their fiducial values: $n_{\rm{H}0} = 10 \, \rm{cm}^{-3}$, $L_{\rm{AGN}} = 10^{46} \, \rm{erg} \, \rm{s}^{-1}$ and $\alpha = 0$. The parameter $\epsilon_{\rm{mix}}$, which determines the fraction of dust grains swept up from the ambient ISM that can mix into the shocked ISM layer after it has cooled (see text), is set to 0.45 (top row) and $10^{-6}$ (bottom row). Once the shocked ISM layer has cooled, dust grains can rapidly re-form via accretion of metals from the gas phase, assuming that dust growth can proceed efficiently in the presence of the strong UV radiation from the AGN). This is true even for very small values of $\epsilon_{\rm{mix}}$, where only a tiny fraction of the dust grains swept up from the ambient ISM are mixed into the shocked ISM layer to seed the accretion process.}
\label{DustFig}
\end{figure} 

In Fig.~\ref{DustFig} we show the evolution of the dust-to-gas ratio in the shocked ISM layer, $\mathcal{D} = M_{\rm{d}} / M_{\rm{s}}$, normalised to the Milky Way value, $\mathcal{D}_{\rm{MW}}$, from our analytic model for different ambient ISM densities (left column) and AGN luminosities (right column) at solar metallicity. The horizontal dotted lines indicate a value of unity. The top row of Fig.~\ref{DustFig} shows the dust evolution for $\epsilon_{\rm{mix}} = 0.45$, which corresponds to the scenario where, once the shocked ISM layer has cooled, all swept up dust grains that initially survive passing through the forward shock are mixed into the shocked ISM layer. At early times, the dust is rapidly destroyed by sputtering, in agreement with \citet{ferrara16}. However, once the shocked ISM layer has cooled, we find that dust can rapidly re-form, due to the accretion of metals from the gas phase onto dust grains. The dust-to-gas ratio continues to rise until it saturates at twice the Milky Way value, where all metals are in dust grains. 

In the bottom row of Fig.~\ref{DustFig}, we show the dust evolution for $\epsilon_{\rm{mix}} = 10^{-6}$. In this scenario, only a very small fraction of swept up dust grains are mixed into the cold phase of the shocked ISM layer after it has cooled. However, we see that this has very little effect on the dust evolution, compared to the top row. Dust grains can thus re-form rapidly via accretion even when only a very small fraction of the dust swept up from the ambient ISM is mixed into the cold phase of the shocked ISM layer to seed the accretion process. Indeed, we see in equation~\ref{acc_time_eqn} that, for typical conditions in the shocked ISM layer after it has cooled, the accretion time-scale ($\approx$$2 \times 10^{4} \, \rm{yr}$) is much shorter than the flow times ($r / v$) of observed outflows ($\sim$10$^{6} \, \rm{yr}$). 

We have therefore shown that our assumption of a constant dust-to-metals ratio in Paper~\textsc{i} is feasible, as dust grains can rapidly re-form via accretion of metals from the gas phase after the shocked ISM layer has cooled, assuming that the grain growth can still be efficient in the presence of the strong UV radiation field of the AGN and assuming that a small ($\ga 10^{-6}$) fraction of dust grains swept up by the outflow after the shocked ISM layer has cooled can be mixed into the cold phase. However, to definitively answer this question of dust survivability in AGN winds and its impact on the formation of molecular outflows, we will need to perform hydrodynamic simulations that model the dust formation and destruction processes in a realistic multiphase AGN wind, coupled to the time-dependent molecular chemistry. 

\section{Conclusions}\label{conclusions_sect} 

In this paper, we have extended the analytic model of FGQ12 for spherically symmetric AGN winds to follow the radiative cooling in the shocked ISM layer of the outflow. We demonstrated in Section~\ref{comparison_sect} that the analytic model reproduces the behaviour of the hydro-chemical AGN wind simulations that we ran in Paper~\textsc{i}. In particular, the analytic model correctly predicts the time at which the shocked ISM layer cools from the post-shock temperature ($\sim 10^{7} \, \rm{K}$) down to $10^{4} \, \rm{K}$ (Fig.~\ref{TSimFig}). We then used the analytic model to explore a wide range of ambient medium densities ($1 \leq n_{\rm{H}0} \leq 10^{4} \, \rm{cm}^{-3}$), density profile slopes ($0 \leq \alpha \leq 1.5$), AGN luminosities ($10^{44} \leq L_{\rm{AGN}} \leq 10^{47} \, \rm{erg} \, \rm{s}^{-1}$), and metallicities ($0.1 \leq Z / \rm{Z}_{\odot} \leq 3$). Our main results are as follows: 

\begin{enumerate} 
\item The time at which the shocked ISM layer cools to $10^{4} \, \rm{K}$ increases with increasing $\alpha$ and $L_{\rm{AGN}}$, and with decreasing $n_{\rm{H}0}$ and $Z$ (Fig.~\ref{TFig}). Apart from the lowest density run ($n_{\rm{H}0} = 1 \, \rm{cm}^{-3}$) and the runs with density slopes $\alpha > 0.5$, all of the analytic models cooled within $1 \, \rm{Myr}$, which corresponds to the typical flow times ($r / v$) of observed molecular outflows in luminous quasars \citep[e.g.][]{gonzalezalfonso17}. Since molecules will form rapidly once the gas has cooled below $10^{4} \, \rm{K}$ (as we showed in Paper~\textsc{i}), we therefore expect molecular outflows to be common across a wide range of physical parameters of AGN winds. 
\item The cooling radius of the outer shell of swept up gas increases with increasing $\alpha$ and $L_{\rm{AGN}}$, and with decreasing $n_{\rm{H}0}$ and $Z$ (Fig.~\ref{rCoolFig}). 
\item The momentum boost factor of the outflow, $\dot{P}_{r} / (L_{\rm{AGN}} / c)$, initially increases as the outflow decelerates, as expected for an energy-conserving flow (Fig.~\ref{PrDotFig}). However, for our fiducial black hole mass ($10^{8} \, \rm{M}_{\odot}$) and host galaxy potential (with velocity dispersion $\sigma = 200 \, \rm{km} \, \rm{s}^{-1}$), the momentum boost factor peaks at $\approx 20$, at an outflow velocity $\approx 200 \, \rm{km} \, \rm{s}^{-1}$, and subsequently declines. This deviation from the simple scaling $\dot{P}_{r}/(L_{\rm{AGN}}/c) \propto 1/v_{\rm s}$ is due to the work done by the outflow against the gravitational potential of the host galaxy. The maximum momentum boost decreases with increasing depth of the gravitational potential (Fig.~\ref{PrDotBHFig}). The momentum boost $-$ outflow velocity relation is insensitive to $n_{\rm{H}0}$, $\alpha$, $L_{\rm{AGN}}$ and $Z$. 
\item Assuming an H$_{2}$ mass fraction $M_{\rm{H}_{2}} / M_{\rm{H}, \, \rm{tot}} = 0.2$ in the shocked ISM layer once it cools below $10^{4} \, \rm{K}$ (based on the results of Paper~\textsc{i}, which assumes the presence of dust grains in the outflow), we find that the H$_{2}$ outflow rate, $\dot{M}_{\rm{H}_{2}}$, at $1 \, \rm{Myr}$ increases slowly with $n_{\rm{H}0}$, by a factor $\approx 10$ as $n_{\rm{H}0}$ increases from 10 to $10^{4} \, \rm{cm}^{-3}$, at $L_{\rm{AGN}} = 10^{47} \, \rm{erg} \, \rm{s}^{-1}$ (Fig.~\ref{MH2DotFig}). $\dot{M}_{\rm{H}_{2}}$ at $1 \, \rm{Myr}$ shows only a weak dependence on $\alpha$, although the high-$\alpha$ models can only form H$_{2}$ within 1 Myr at low AGN luminosities. The intermediate-density analytic models ($n_{\rm{H}0} = 10 - 100 \, \rm{cm}^{-3}$) agree well with the observed $\dot{M}_{\rm{H}_{2}}$ of \citet{cicone14} if we use the CO to H$_{2}$ conversion factors, $\alpha_{\rm{CO}}$, predicted by the simulations in Paper~\textsc{i}. The outflow velocities from the analytic model generally underpredict the maximum velocities of observed outflows. This is likely due at least in part to the lack of ambient inhomogeneities in the analytic model, which results in a single outflow velocity rather than a distribution of velocities, as found in observations.
\item If we consider an observational estimate for the momentum boost factor, $v_{\rm{s}} \dot{M}_{\rm{H}_{2}} / (L_{\rm{AGN}} / c)$, which is averaged over the flow time of the outflow, rather than the instantaneous rate of change of radial momentum (which is often used to quantify the momentum boost in theoretical models), the analytic models predict a maximum momentum boost of $\approx 1 - 2$ (Fig.~\ref{PrDotObsFig}). This is so even for energy-conserving models for which the boost measured in terms of the instantaneous rate of change of the radial momentum is instead up to $\approx 20$. This is partly due to our conservative estimate for the H$_{2}$ fraction of 0.2. However, we also show that the observational estimate is not equivalent to the instantaneous time derivative of the radial momentum. We therefore conclude that, while recent observations of AGN winds have estimated momentum boosts of order unity \citep[e.g.][]{feruglio17, veilleux17}, these do not necessarily rule out an energy-driven outflow. 
\item By modelling the formation and destruction of dust grains in the shocked ISM layer in our analytic AGN wind model, we find that dust grains can rapidly re-form in the wind via accretion of metals from the gas phase after the shocked ISM layer has cooled. This is true even when the fraction of dust grains swept up after the shocked ISM layer has cooled that survive and mix into this layer to seed the accretion process is very small ($\sim$10$^{-6}$). This results in a high dust-to-gas ratio (close to the Milky Way value) that enables molecule formation in the AGN wind. However, we caution that the accretion time-scale used in this dust model is uncertain, as it assumes a sticking coefficient of metals on to dust grains of unity, even at gas temperatures $\sim$10$^{4}$ K corresponding to the shocked ISM layer after it has cooled in our model, and it neglects the effects of grain charging by the strong UV radiation field from the AGN, which may further suppress dust growth. 
\end{enumerate} 

We have thus demonstrated that molecular outflows can potentially form across a wide range of physical AGN wind parameters. The results of this paper allow us to extend the predictions of our molecular outflow simulations from Paper~\textsc{i} to a wider range of physical parameters than with the simulations alone. By comparing these predictions to observations, we can then test these models, and the assumptions that go into them. In particular, we will show in a forthcoming paper that the strong mid-infrared emission from warm H$_{2}$ (at a few hundred K) that we found in Paper~\textsc{i} will be detectable with the \textit{James Webb Space Telescope (JWST)} at high signal to noise ratios, which will enable the emission from the outflow to be spatially and kinematically distinguished from the host galaxy. Future \textit{JWST} observations of this warm H$_{2}$ emission will thus be a key test for our models. Additionally, these results can be used to guide future simulations of molecule formation in AGN winds, as well as to enable more accurate comparisons of energy-conserving wind models to observations. 

\section*{Acknowledgements}

We thank the referee, Evan Scannapieco, for his detailed report, which improved the quality of this manuscript. We also thank Eduardo Gonz\'{a}lez-Alfonso for his detailed comments, as well as Eliot Quataert, Paul Torrey, and Phil Hopkins for useful discussions. AJR is supported by the Lindheimer fellowship at Northwestern University. CAFG was supported by NSF through grants AST-1412836, AST-1517491, AST-1715216, and CAREER award AST-1652522, by NASA through grant NNX15AB22G, by CXO through grant TM7-18007X, and by a Cottrell Scholar Award from the Research Corporation for Science Advancement. The simulations used in this work were run on the Stampede supercomputer at the Texas Advanced Computing Center (TACC) through allocations TG-AST160035 and TGAST160059 granted by the Extreme Science and Engineering Discovery Environment (XSEDE), which is supported by NSF grant number ACI-154562; the Pleiades supercomputer through allocation s1480, provided through the NASA Advanced Supercomputing (NAS) Division at Ames Research Center; and the Quest computing cluster at Northwestern University, which is jointly supported by the Office of the Provost, the Office for Research, and Northwestern University Information Technology.

{}

\appendix 

\section{Resolution tests}\label{resolution_appendix} 

The fiducial resolution of the simulations from Paper~\textsc{i} was $30 \, \rm{M}_{\odot}$ per gas particle, with a minimum gravitational softening for gas particles of $0.1 \, \rm{pc}$. In Paper~\textsc{i}, we also repeated these simulations with a factor 8 lower mass resolution, and the low-luminosity run with a factor 3 higher mass resolution. In this section, we use these low- and high-resolution runs to test the numerical convergence of the simulation results presented in Section~\ref{comparison_sect}. 

\begin{figure}
\centering
\mbox{
	\includegraphics[width=84mm]{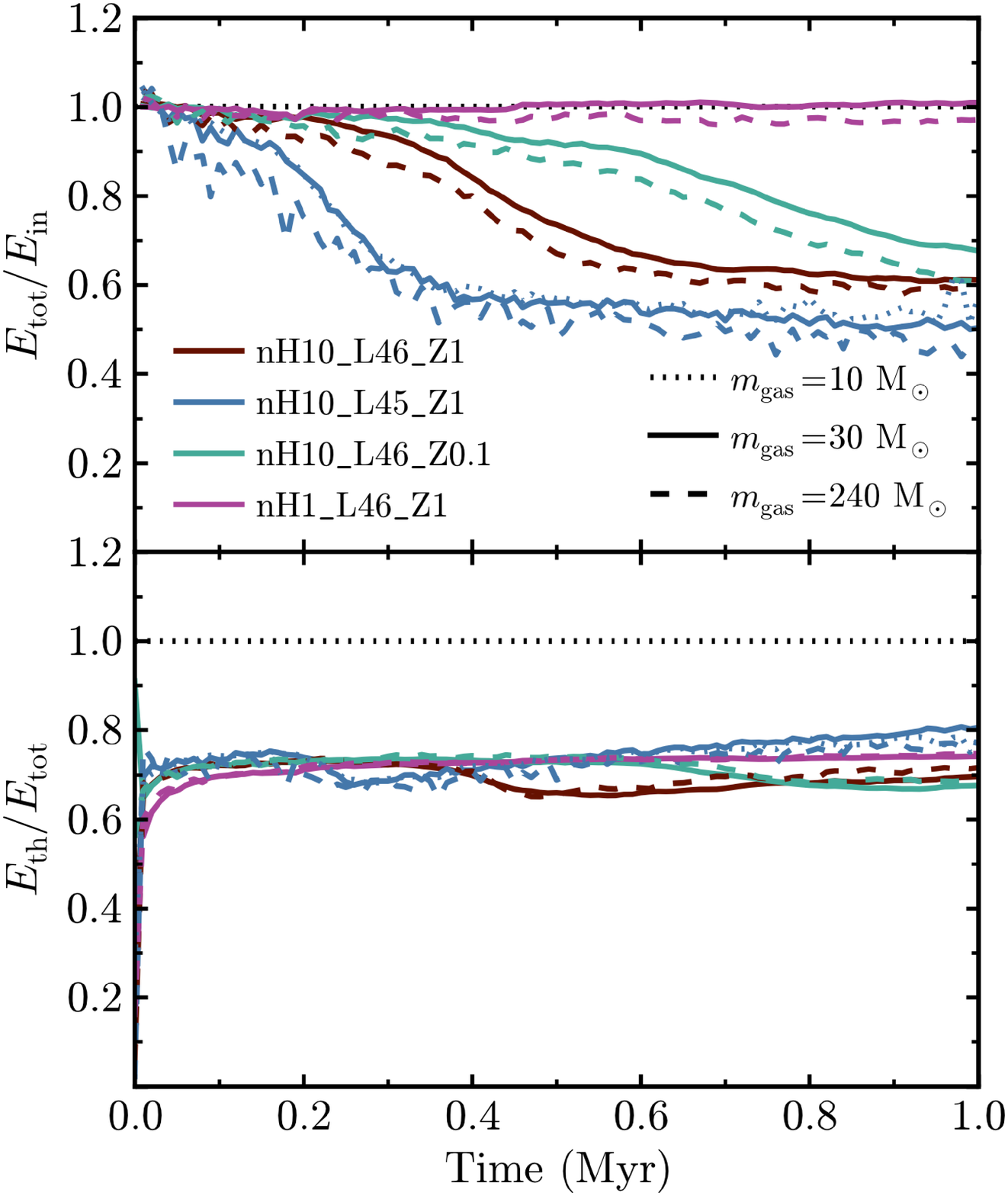}}
\caption{Ratio of the total (thermal plus kinetic) energy of the outflow to the mechanical wind energy injected by the AGN ($E_{\rm{tot}} / E_{\rm{in}}$, top panel), and the fraction of the outflow energy that is thermal ($E_{\rm{th}} / E_{\rm{tot}}$, bottom panel), plotted against time, for runs nH10\_L46\_Z1 (red curves), nH10\_L45\_Z1 (blue curves), nH10\_L46\_Z0.1 (green curves), and nH1\_L46\_Z1 (magenta curves), at low- (dashed curves), fiducial (solid curves), and high-resolution (dotted curves). We see that these ratios are well converged in the simulations.} 
\label{energyResTestFig}
\end{figure}

The top panel of Fig.~\ref{energyResTestFig} shows the time evolution of the ratio of the total (thermal plus kinetic) outflow energy ($E_{\rm{tot}}$) to the energy injected by the AGN ($E_{\rm{in}}$) for the simulations nH10\_L46\_Z1 (red curves), nH10\_L45\_Z1 (blue curves), nH10\_L46\_Z0.1 (green curves), and nH1\_L46\_Z1 (magenta curves), at low- (dashed curves), fiducial (solid curves), and high-resolution (dotted curves). In the bottom panel, we show the time evolution of the fraction of the outflow energy that is thermal in these runs. We see that these energy ratios are in very good agreement at different resolution levels. 

\begin{figure}
\centering
\mbox{
	\includegraphics[width=84mm]{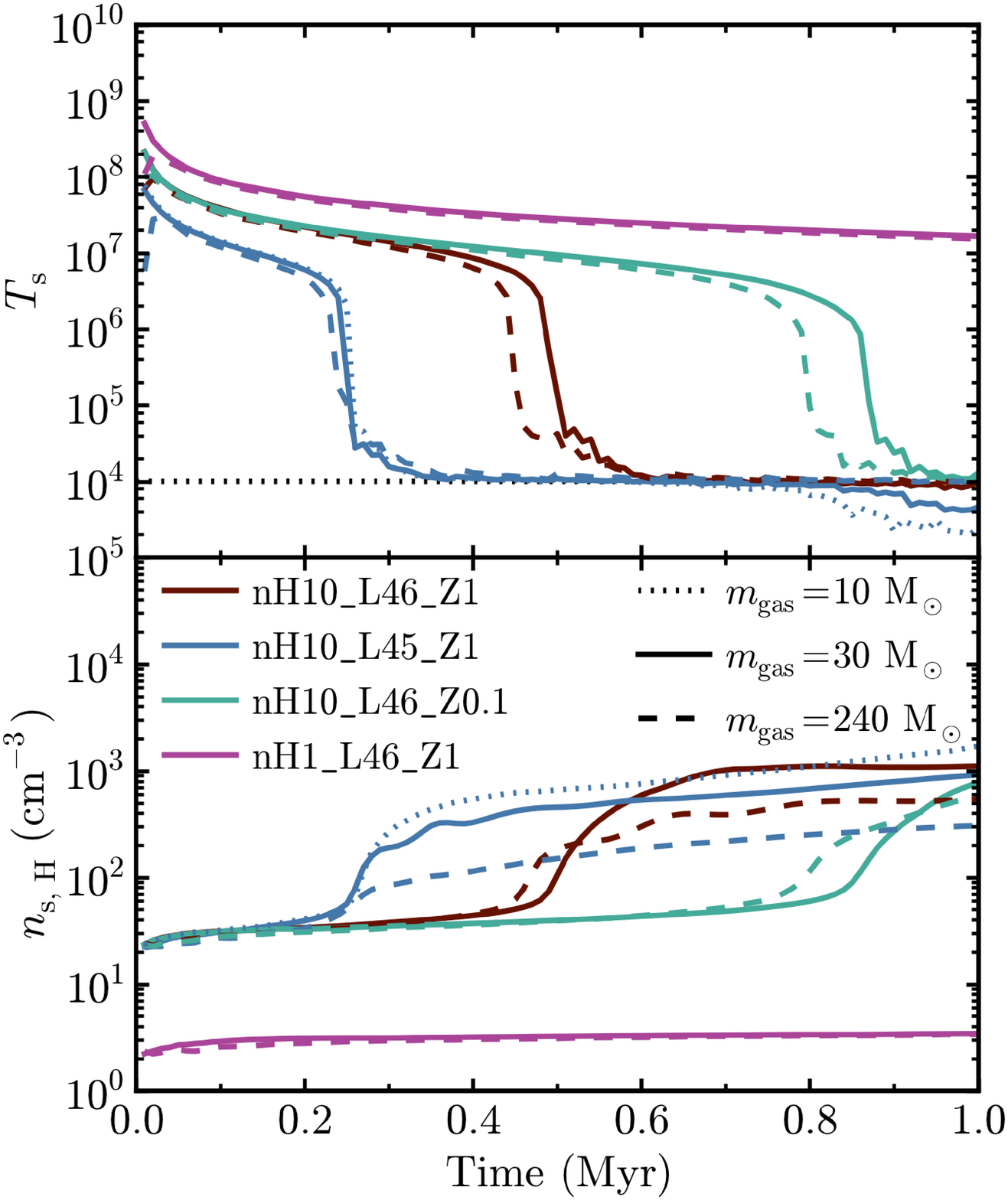}}
\caption{Time evolution of the median temperature ($T_{\rm{s}}$, top panel) and hydrogen density ($n_{\rm{s}, \, \rm{H}}$, bottom panel) of the shocked ISM layer (defined as particles with densities $>2 n_{\rm{H}0}$) in the simulations at different resolution levels. The horizontal dotted line in the top panel indicates a temperature of $10^{4} \, \rm{K}$, at which the cooling function has been truncated. In the fiducial and low-metallicity runs (red and green curves, respectively), the shocked ISM layer cools $\approx 10$ per cent sooner at low-resolution (dashed curves) than at standard resolution (solid curves), although the temperature evolution of the low-luminosity run is well converged. After the shocked ISM layer has cooled, there is also a trend of increasing median density with increasing resolution, which is unsurprising as higher resolutions can resolve gas to higher densities.} 
\label{temperatureResTestFig}
\end{figure}

Fig.~\ref{temperatureResTestFig} shows the time-evolution of the median temperature ($T_{\rm{s}}$, top panel) and hydrogen density ($n_{\rm{s}, \, \rm{H}}$, bottom panel) of the shocked ISM layer, which we define as particles with densities $>2 n_{\rm{H}0}$. In the fiducial and low-metallicity simulations (red and green curves, respectively), we see that the rapid drop in $T_{\rm{s}}$ begins $\approx 10$ per cent sooner at low-resolution (dashed curves) than at standard resolution (solid curves), although the time at which they reach $10^{4} \, \rm{K}$ (the horizontal dotted line) is in good agreement at different resolutions. The temperature evolution of the low-luminosity run (blue curves) is well converged, while the low-density run (magenta curves) does not cool within $1 \, \rm{Myr}$. 

Once the shocked ISM layer has cooled, there is a trend of increasing median density in the shocked ISM layer with increasing resolution. This is unsurprising because, at higher resolution, we are able to resolve gas structures up to higher densities. 

\begin{figure}
\centering
\mbox{
	\includegraphics[width=84mm]{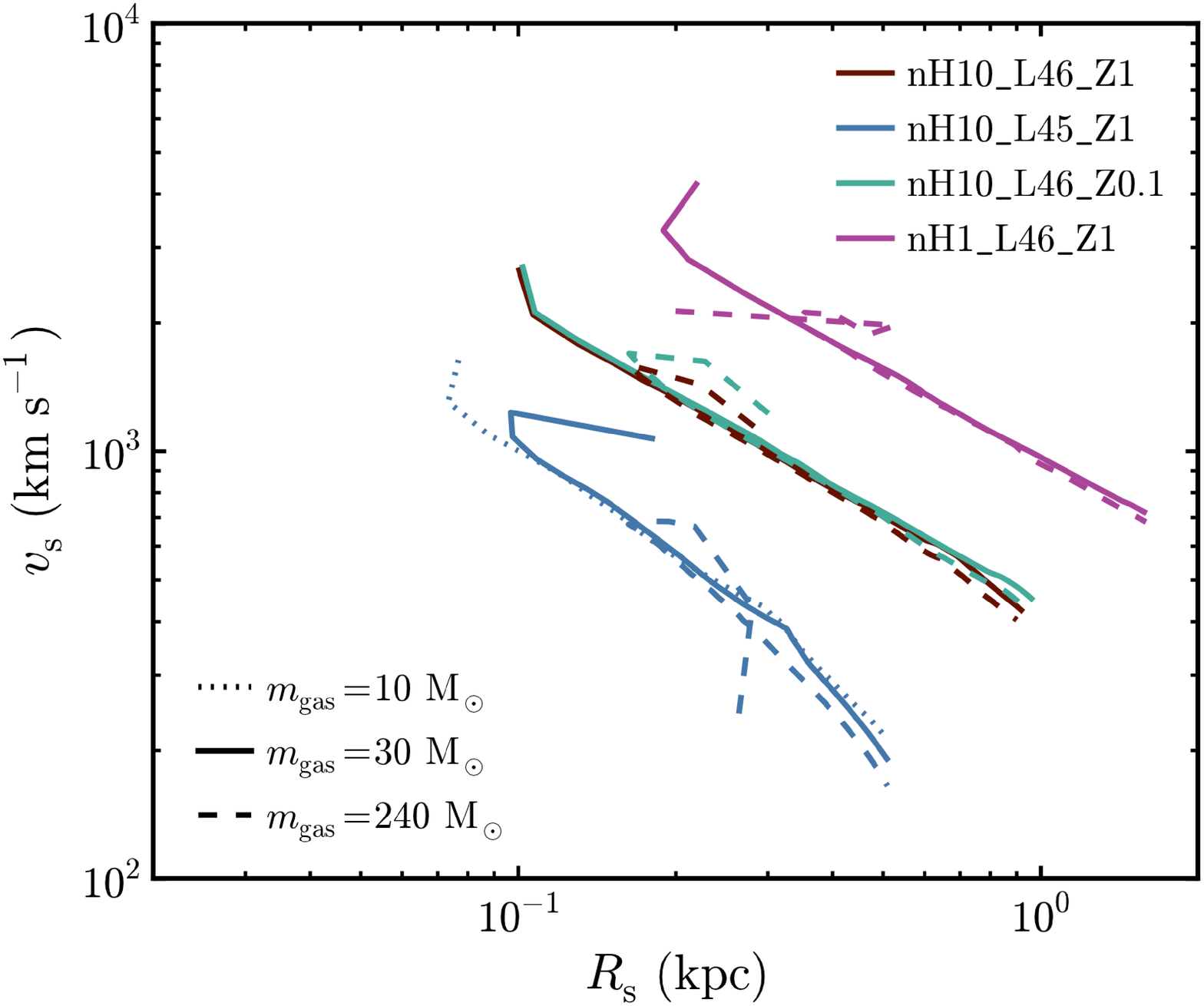}}
\caption{Mass-weighted mean velocity ($v_{\rm{s}}$) versus mass-weighted mean radius ($R_{\rm{s}}$) of particles in the shocked ISM layer (with densities $>2 n_{\rm{H}0}$) from the simulations at different resolution levels. At early times, the radius and velocity of the shocked ISM layer is poorly defined, because it takes a finite time for this shell to build up to the post-shock density of $4 n_{\rm{H}0}$, which is needed by our definition of the shocked ISM layer based on a density cut. This becomes more problematic at lower resolution, which explains the discrepant behaviour of the low-resolution runs (dashed curves) in this plot. However, at late times, once the shocked ISM layer is well established, the simulations at different resolutions are very well converged.} 
\label{vsRsResTestFig}
\end{figure}

Fig.~\ref{vsRsResTestFig} shows the mass-weighted mean velocity of particles in the shocked ISM layer ($v_{\rm{s}}$), plotted against their mean radius ($R_{\rm{s}}$), from the simulations at different resolution levels. At early times, the shocked ISM layer is poorly defined, because it takes a finite time for this layer to build up to the expected post-shock density of $4 n_{\rm{H}0}$. This is important as we define the shocked ISM layer based on a density cut ($>2 n_{\rm{H}0}$). This is especially problematic at low resolution, where it takes longer for the shocked ISM layer to become well defined, which leads to the discrepant behaviour of the low-resolution runs (dashed curves) in this figure. However, at late times, once this layer is well defined, we see that the $R_{\rm{s}} - v_{\rm{s}}$ relation is well converged with resolution. 

\begin{figure}
\centering
\mbox{
	\includegraphics[width=84mm]{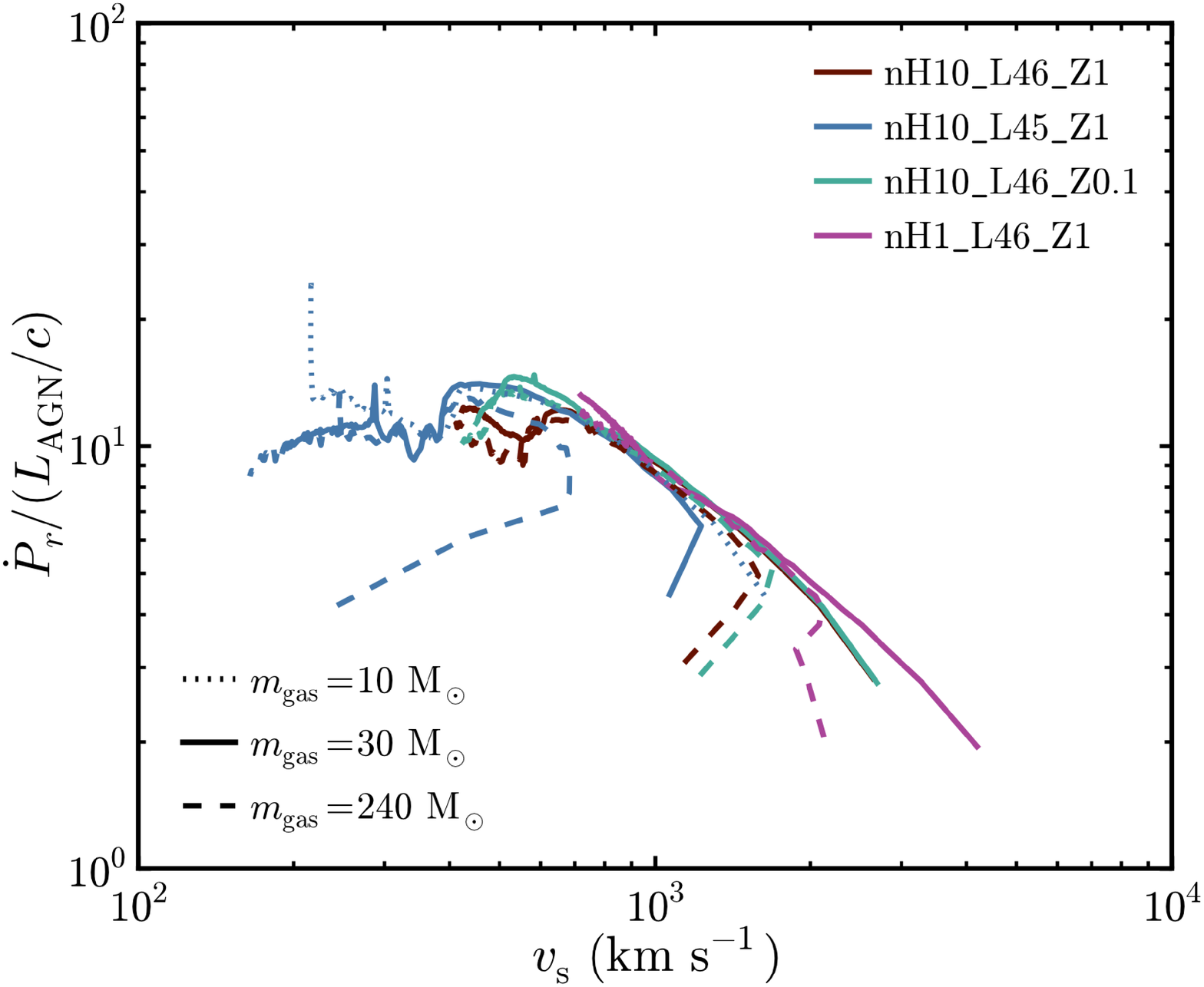}}
\caption{Rate of change of radial momentum of the entire outflow (including the hot shocked wind bubble) relative to the momentum injection rate of the AGN ($\dot{P}_{r} / (L_{\rm{AGN}} / c)$), plotted against the mass-weighted mean velocity ($v_{\rm{s}}$) of particles in the shocked ISM layer, from simulations at different resolution levels. As in Fig.~\ref{vsRsResTestFig}, the shocked ISM layer is poorly defined at early times (which corresponds to the right-hand side of the plot, as the outflow is decelerating), especially in the low-resolution runs, which leads to the discrepant behaviour of the low-resolution runs (dashed curves). However, once the shocked ISM layer is well defined, the different resolution levels are mostly well-converged. In particular, the maximum momentum boost is $\approx 10$ in all runs, except in the high-resolution run of nH10\_L45\_Z1, which shows an anomalous peak in the final snapshot.} 
\label{PrDotResTestFig}
\end{figure}

In Fig.~\ref{PrDotResTestFig} we show the momentum boost factor, defined as the rate of change of radial momentum of the outflow relative to the momentum injection rate of the AGN ($\dot{P}_{r} / (L_{\rm{AGN}} / c)$), plotted against the mean velocity of the shocked ISM layer, from the simulations at different resolutions. As noted above, the shocked ISM layer is poorly defined at early times, especially at low resolution. This leads to the discrepant behaviour of the low-resolution runs (dashed curves) at early times, which corresponds to the right-hand side of this plot, since the outflow is decelerating, so the outflow moves from right to left in this plot. However, once the the shocked ISM layer is well defined, we generally find good numerical convergence in the momentum boost factor. In particular, the maximum momentum boost is $\approx 10$ in all runs, independent of resolution, except for the high-resolution run of the low-luminosity simulation (blue dotted curve), which shows an anomalous peak in the final snapshot. 

\section{Inverse Compton cooling of the shocked AGN wind}\label{2T_cool_sect} 
At the very high temperatures ($T_{\rm sh} \ga 10^{9}$ K) and low densities of shocked AGN winds driven by accretion disks with initial velocities $v_{\rm in} \ga 10 \, 000$ km s$^{-1}$, the dominant radiative cooling mechanism is inverse Compton scattering \citep[e.g.][]{king03}. Since the Compton temperature of the AGN radiation field $T_{\rm C}\sim 10^{7}$ K, Compton scattering of AGN photons with electrons in the shocked wind takes thermal energy away from the shocked wind. FGQ12 noted two subtleties in accurately evaluating the effective cooling rate of the shocked wind in these conditions. 

First, $T_{\rm sh} \sim 10^{9}$ K is right around the transition between non-relativistic and relativistic electrons. This is important because the inverse Compton cooling rate is independent of the electron temperature, $T_{\rm e}$, in the non-relativistic regime, but is $\propto T_{\rm e}$ when the electrons become relativistic. For very hot, relativistic electrons, the inverse Compton cooling time becomes short and this can cause the wind bubble to cool rapidly. The inverse Compton cooling time remains longer when the electrons remain non-relativistic. This helps wind bubbles retain their thermal energy and stay in the energy-conserving limit. Second, to determine when the shocked wind bubble loses its thermal pressure support, we must evaluate the cooling rate of the protons. Since $m_{\rm p} \gg m_{\rm e}$ and $\sigma_{\rm T} \propto m_{i}^{-2}$ (where $\sigma_{\rm T}$ is the Thomson scattering cross section), protons do not directly lose significant energy via inverse Compton scattering. Rather, they cool through interactions with electrons. 

FGQ12 argued that, regardless of how electrons are heated by electromagnetic turbulence at the shock, if collective effects die down within a reasonable distance/time past the shock, then a two-temperature (2T) plasma should develop with equilibrium electron temperature $T_{\rm e}^{\rm eq} < T_{\rm p}$. This equilibrium temperature is determined by a balance between inverse Compton cooling and heating of the electrons by Coulomb collisions with protons. When this equilibrium is reached, the cooling rate of the protons is the rate at which they transfer energy to the electrons via Coulomb collisions, which is equal to the rate at which electrons cool via inverse Compton scattering.

FGQ12 showed how to model the effects of 2T cooling in spherically-symmetric AGN wind calculations. In this appendix, we derive a more general prescription that can be used to approximate the effects of 2T cooling in hydrodynamic simulations of AGN winds, including in 3D. 

The volumetric cooling rate, $\Lambda_{\rm{IC}, \, \rm{2T}}$, of protons due to Coulomb collisions with electrons in the shocked wind is given by:  
\begin{align}
\Lambda_{\rm{IC}, \, \rm{2T}} = \frac{3 k_{B} n_{\rm p}}{2} \frac{dT_{\rm p}}{dt}
\end{align}
In a fully ionized, neutral plasma consisting of free protons and free electrons interacting solely via Coulomb collisions, the proton temperature evolves following
\begin{align}
\label{eq:dTpdt}
\frac{dT_{\rm p}}{dt} = \frac{T_{\rm e}-T_{\rm p}}{t_{\rm ei}},
\end{align}
where the Coulomb equilibration time
\begin{align}
\label{t_ei}
t_{\rm ei} & = \frac{3 m_{\rm e} m_{\rm p}}{8(2 \pi)^{1/2} n_{\rm p} e^{4} \ln{\Lambda}} \left( \frac{k_{B} T_{\rm e}}{m_{\rm e}} + \frac{k_{B} T_{\rm p}}{m_{\rm p}} \right)^{3/2}.
\end{align}
In this expression, the Coulomb logarithm (not to be confused with the logarithm of the cooling rate) is given by
\begin{equation}
\label{eq:Coloumb_log}
\ln{\Lambda} \approx 39 + \ln{\left( \frac{T_{\rm e}}{\rm 10^{10}~K} \right)} - \frac{1}{2} \ln{\left( \frac{n_{\rm e}}{\rm 1~cm^{-3}} \right)} 
\end{equation}
\citep{spitzer62} and $e$ is the electric charge of the electron. Following FGQ12, we are interested in the regime $T_{\rm e} \sim 0.1T_{\rm p}$. In this case, $T_{\rm e} \ll T_{\rm p}$ but $T_{\rm e}/m_{e} \gg T_{\rm p}/m_{p}$. Equation (\ref{eq:dTpdt}) then simplifies to $dT_{\rm p}/dt = -T_{\rm p} / t_{\rm ei}$ and equation (\ref{t_ei}) simplifies to $t_{\rm ei} \propto (k_{B} T_{\rm e}/m_{\rm e})^{3/2}$. Combining,
\begin{align}
\Lambda_{\rm{IC}, \, \rm{2T}} &= - \frac{4 (2 \pi)^{1/2} n_{\rm p}^{2} e^{4} m_{\rm e}^{1/2} \ln{\Lambda}}{\beta^{3/2} m_{\rm p} (k_{B} T_{\rm p})^{1/2}} \nonumber \\ 
& \approx 1.0\times10^{-19}~{\rm erg~cm^{-3}~s^{-1}} \left(\frac{\beta}{0.1}\right)^{-3/2} \left(\frac{T_{\rm p}}{\rm 10^{8}~K}\right)^{-1/2} \nonumber \\ 
& \hspace{1.3 in} \times \left(\frac{n_{\rm{p}}}{1 \, \rm{cm}^{-3}}\right)^{2} \left(\frac{\ln{\Lambda}}{40}\right),
\end{align}
where $\beta \equiv T_{\rm e} / T_{\rm p}$. 

The value $\beta \approx 0.1$ is representative of the spherically-symmetric wind solutions presented in FGQ12. However, a general self-consistent application requires evaluating how $\beta$ depends on local physical conditions. Under assumptions consistent with those above, FGQ12 showed that for $T_{\rm e}=T_{\rm e}^{\rm eq}$, 
\begin{align}
\label{Te eq}
\beta = \frac{T_{\rm e}^{\rm eq}}{T_{\rm p}} & \approx \frac{(2 \pi)^{1/5}}{T_{\rm p}^{3/5}} 
\left[
\frac{m_{\rm e}^{3} e^{8} c^{2} (\ln{\Lambda})^{2} n_{\rm p}^{2}}{\sigma_{\rm T}^{2} k_{B}^{3} m_{\rm p}^{2} U_{\rm ph}^{2}}
\right]^{1/5},
\end{align}
where $U_{\rm ph}$ is the energy density in the radiation field. FGQ12 derived this analytic expression for $T_{\rm e}^{\rm eq}$ in the limit in which inverse Compton cooling is well approximated by the expression for non-relativistic electrons, which they found is generally applicable for shocked AGN winds because 2T effects keep the electrons cooler than the protons. This gives
\begin{align}
\Lambda_{\rm{IC}, \, \rm{2T}}(T_{\rm e}=T_{\rm e}^{\rm eq}, n_{\rm{p}}) = 
4 \left[
\frac{2 \pi k_{B}^{2} \sigma_{\rm T}^{3} e^{8} (\ln{\Lambda})^{2} T_{\rm p}^{2} n_{\rm p}^{7} U_{\rm ph}^{3}}{c^{3} m_{\rm p}^{2} m_{\rm e}^{2}}
\right]^{1/5}.
\end{align}
By construction, $\Lambda_{\rm{IC}, \, \rm{2T}}(T_{\rm e}=T_{\rm e}^{\rm eq}) = \Lambda_{\rm IC}(T_{\rm e}=T_{\rm e}^{\rm eq})$, where $\Lambda_{\rm IC}$ is the standard inverse Compton cooling rate, since $T_{\rm e}^{\rm eq}$ is defined such that the Compton cooling rate of the electrons equals the Coulomb heating rate by protons. Since the Coulomb logarithm depends only logarithmically on temperature (equation~\ref{eq:Coloumb_log}), it can be evaluated using the proton temperature instead of the electron temperature without introducing a large error.

We now synthesize the above results into a general prescription:
\begin{equation}\label{lambda_2T_eqn} 
  \Lambda_{\rm{IC}, \, \rm{2T}}(T_{\rm p}, n_{\rm{p}}) = 
  \begin{cases} 
    \Lambda_{\rm IC}(T_{\rm e}=T_{\rm e}^{\rm eq}) & 10T_{\rm C} < T_{\rm e}^{\rm eq} \leq T_{\rm p} \\ 
    \Lambda_{\rm IC}(T_{\rm e}=T_{\rm p}) & \rm{otherwise}. 
  \end{cases} 
\end{equation}
In the above equation, $T_{\rm e}^{\rm eq}=T_{\rm e}^{\rm eq}(T_{\rm p},~n_{\rm p},~U_{\rm ph})$ and we identify the proton temperature with the temperature of the gas tracked by the hydrodynamics solver. The first conditional ($10T_{\rm C} < T_{\rm e}^{\rm{eq}}$) is included because of the assumption that $T_{\rm e} \gg T_{\rm C}$ in the above derivation. The second conditional ($T_{\rm e}^{\rm eq} \leq T_{\rm p}$) should always be realized in conditions representative of shocked AGN winds because inverse Compton cooling should keep the electrons cooler than the protons; it is included only as a limiter to avoid potentially pathological behavior. This prescription for $\Lambda_{\rm{IC}, \, \rm{2T}}$ reduces to ordinary inverse Compton cooling for low-temperature, single-temperature plasmas. To avoid double counting inverse Compton cooling, simulations should replace the usual inverse Compton cooling rate with $\Lambda_{\rm{IC}, \, \rm{2T}}$.

\label{lastpage}

\end{document}